\documentclass[showpacs,amsmath,amssymb,10pt,aps]{revtex4}

\newtheorem{Proposition}{Proposition}
\newtheorem{theorem}{Theorem}
\newtheorem{Remark}{Remark}

\usepackage{bbm}
\usepackage{bm}
\usepackage{physics}
\usepackage[T1]{fontenc}
\usepackage[cp1250]{inputenc}
\usepackage{amsbsy}

\begin{document}

\title{Stochastic approach to evolution of a quantum system interacting with a wave packet in squeezed number state}

\author{Anita D\k{a}browska}
\author{Marcin Marciniak}

\affiliation{Institute of Theoretical Physics and Astrophysics,  University of Gda\'nsk, ul. Wita Stwosza 57, Gda\'nsk, 80-308, Poland}


\begin{abstract}
	We determine filtering and master equations for a quantum system interacting with wave packet of light in a continuous-mode squeezed number state. We formulate the problem of conditional evolution of a quantum system making use of model of repeated interactions and measurements. In this approach the quantum system undergoes a sequence of interactions with an environment defined by a chain of harmonic oscillators. We assume that the environment is prepared in an entangled state being a discrete analogue of a continuous-mode number state. We present a derivation of a discrete stochastic dynamics that depends on the results of measurement performed on the field after its interaction with the system. In this paper we consider a photon counting measurement scheme. By taking a continuous time limit, we finally obtain differential stochastic equations for the system. Analytical formulae for quantum trajectories and exclusive probability densities that allow to fully characterize the statistics of photons in the output field are given.
\end{abstract}    	 


\maketitle

\section{Introduction}

Quantum filtering theory \cite{BarBel91,Car93,Bar06,GZ10,WM10}, formulated in the framework of quantum Ito stochastic calculus (QSC)  \cite{HP84,Par92}, provides mathematical tools that are used to study the evolution of an open quantum system conditioned by results of a continuous in time measurement performed on a field that has interacted with the system. 
The evolution of the composed system (quantum system and the field) is given there by unitary operator which satisfies quantum stochastic differential equation of It\^{o} type  \cite{HP84,Par92}. The filtering theory is connected to the input-output formalism \cite{GZ10}. Information about the system, carried by the output field, can be gained in different detection schemes: direct photon counting, homodyne, or heterodyne measurement. The random results of the measurement give rise to stochastic evolution of the open system. The conditional evolution of the system is described by equation that is known as filtering or stochastic master equation (SME). Its form depends on the state of the input field and the type of detection that are taken. Solutions to the filtering equation are called {\it a posteriori} states or quantum trajectories. If the measurement is not performed or its results are not taken into account then the evolution of the open system is given by the master equation.

The methods of quantum filtering theory were originally developed for the input field prepared in a Gaussian state (vacuum, coherent, squeezed, thermal). In that case the unconditional evolution of the system is given by a single master equation \cite{BarBel91,Car93,Bar06,GZ10,WM10} of the Lindbland form \cite{GKS76,Lin76}. The recent development of experimental techniques of producing and manipulating propagating wave packets of light in non-classical states \cite{CWSS13,PSZ13,RR15,Leong2016,Zoller2017}, such as $n$-photon state, squeezed vacuum state \cite{BLPS90,Loudon2000,O06,RMS07}, and application of such packets, for instance, in quantum communication and quantum cryptography \cite{Scarani09,Aaronon11,Sedziak2017,Zhong2022} brought the extension of the filtering theory methods. The wave packet of the $n$-photon state possesses temporal correlations that make the dynamics of the system non-Markovian \cite{Breuer2016,Ciccarello2018,Dabrowska2021}. The reduced evolution of the system is no longer given then by a single equation but by a set of coupled equations \cite{   Gheri98,Baragiola12,Gough12a,Gough12b,Gough13, Dong15,Song16,Pan16,Dabrowska17,Baragiola17,Dabrowska18,Dabrowska19,Dabrowska2020}. To study the quantum trajectories for the system interacting with non-classical states several methods were proposed. One can use the cascaded system approach \cite{Gough12a,Dabrowska18}, a non-Markovian embedding technique \cite{Gough12b,Gough13,Song16}, or a temporal decomposition of the input field \cite{Baragiola17,Gross2022}. All mentioned approaches were formulated by means of QSC. There exists also a method based on a discrete model of repeated interactions and measurements, see \cite{Dabrowska17,Dabrowska18,Dabrowska19}. 

In this paper, we study the interaction of a quantum system with a traveling light prepared in a continuous-mode squeezed number state \cite{Loudon2000}. This problem was recently analyzed by Gross {\it et. al.} in the framework of QSC in \cite{Gross2022}. 
We consider the electromagnetic field of a finite time correlation. In the literature there exist the filtering equations formulated for the broadband squeezing bath \cite{GZ10,WM10,Gough2003,Bouten2004,Dabrowska2016a,Dabrowska2016b}. 
Such field, called also the squeezed white-noise field, has $\delta$-time correlations and the infinite photon flux. Due to this property the stochastic equation for the broadband squeezing bath for photon counting measurement can not be defined. The interaction of the quantum system with a finite and narrow-bandwidth squeezed field were studied, for instance in \cite{Carmichael1973,Yeoman1996,Ficek1997,Tanas1999,Kowalewska2001,Parkins1990,Messikh2000}. 

We define the interaction of open quantum system with the light making use of the model of repeated interactions and measurements, called also a collision model \cite{B02,GS04,AP06,P08,PP09,P10,BHJ09,Kretschmer2016,Ciccarello2017,ACMZ17,Filippov2022,Gross18,Ciccarello2021}. 
A detailed discussion on physical assumptions for the collision models in quantum optics was given, for instance, in \cite{Gross18,Ciccarello2017,Fischer2018a}. The main aim of our paper is to show that the discrete approach provides a rigorous and intuitive way for studying quantum conditioning and stochastic evolution for a quantum system coupled to the field in the squeezed number state. We define the environment (a traveling unidirectional field) by a chain of harmonic oscillator that is initially prepared in a state that is discrete analog of continuous-mode squeezed number state. We assume that the quantum system undergoes a sequence of interactions  ("collisions") with elements of the environment. There are no initial correlations between the environment and the system. 
The harmonics oscillators do not interact with each other but they interact with the system one by one and they are subsequently monitored. Random results of the measurements lead to random sequence of the system states. We derive discrete in time set of stochastic recurrence equations describing the conditional evolution of the system and we display the analytical solution related to different realization of the stochastic process connected to the measurement. Finally, we obtain the continuous in time conditional and unconditional evolution of the quantum system. The set of filtering and master equations derived in this papers agrees with the results given in \cite{Gross2022}. We not only derive differential equations for conditional and unconditional evolution of the system, but also determine the analytical formulae for the quantum trajectories related to the continuous in time counting detection of the output field. And this allows us to characterize the whole statistics of the output photons by the exclusive probability densities \cite{Srinivas81,Mollow68}. Finally, we show how to use conditional vectors to determine a photon profile for an optimal transfer of photons from the wave packet into the cavity.

The paper has the following structure. Section 2 is devoted to presentation of the model of repeated interactions. 
In Section 3 we define the squeezed number state for the environment and describe some of its properties. In Section 4 we give a description of quantum conditioning for the sequence of repeated measurements.   
Section 4 is devoted to determination of the sets of discrete filtering and master equations and their limit for the continuous in time evolution of the system. In Section 5 we present the general analytical formulae for quantum trajectories and use it to define the statistics of the output photons. In Section 6 we give an example of usage of the conditional operators to solve the problem of efficient transfer of photons. We summarize our results in Section 7. 

\section{Repeated interaction model}

We consider a quantum system $\mathcal{S}$, described by the Hilbert space $\mathcal{H_{S}}$, interacting with an environment (a unidirectional  field) $\mathcal{E}$ defined by a sequence of $M$ harmonic oscillators. We assume that the field harmonic oscilators do not interact with each other but they interact one by one with the system $\mathcal{S}$ each during a time $\tau$. 
The Hilbert space of the environment is given then by
\begin{equation}
\mathcal{H}_{\mathcal{E}}=\bigotimes_{k=0}^{M-1}\mathcal{H}_{\mathcal{E},k},
\end{equation}
where $\mathcal{H}_{\mathcal{E},k}$ is the Hilbert space of the harmonic oscillator which interacts with $\mathcal{S}$ in the time interval $[k\tau, (k+1)\tau)$. Clearly, $\mathcal{H}_{\mathcal{E}}$ posses the following property: for each $j\geq 1$ we have
\begin{equation}
\mathcal{H}_{\mathcal{E}}=\mathcal{H}_{\mathcal{E}}^{j-1]}\otimes \mathcal{H}_{\mathcal{E}}^{[j},
\end{equation}
where
\begin{equation}
\mathcal{H}_{\mathcal{E}}^{j-1]}= \bigotimes_{k=0}^{j-1}\mathcal{H}_{\mathcal{E},k},\;\; \mathcal{H}_{\mathcal{E}}^{[j}=\bigotimes_{k=j}^{M-1}\mathcal{H}_{\mathcal{E},k}.
\end{equation}
We can split the Hilbert space $\mathcal{H}_{\mathcal{E}}$ into the input and output parts referring respectively to the harmonic oscillators which have already interacted with $\mathcal{S}$ and those which will interacts with $\mathcal{S}$ in the future. The total Hilbert space is $\mathcal{H}_{\mathcal{E}}\otimes \mathcal{H_{S}}$. We describe the dynamics of the composed system $\mathcal{E}+\mathcal{S}$ up to time $T=M\tau$. At a given moment $\mathcal{S}$ interacts with only one of the field harmonic oscillators. An interaction of the field with $\mathcal{S}$ in time interval $[k\tau, (k+1)\tau)$
is defined by the unitary operator:
\begin{equation}\label{intermat}
\hat{\mathbb{V}}_{k}= \hat{\mathbbm{1}}_\mathcal{E}^{k-1]} \otimes \hat{\mathbb{V}}_{[k}  ,
\end{equation}
where
\begin{equation}\label{Vk}
\hat{\mathbb{V}}_{[k} =  \exp\left(-i\tau \hat{H}_{k}\right)
\end{equation}
and
\begin{eqnarray}\label{hamint}
\hat{H}_{k} = \hat{\mathbbm{1}}_{\mathcal{E}}^{[k}\otimes \hat{H}_{\mathcal{S}}+\frac{i}{\sqrt{\tau}}\left(\hat{b}_{k}^{\dagger}\otimes\hat{\mathbbm{1}}_{\mathcal{E}}^{[k+1}\otimes \hat{L}-\hat{b}_{k}\otimes\hat{\mathbbm{1}}_{\mathcal{E}}^{[k+1}\otimes \hat{L}^{\dagger}\right).
\end{eqnarray}
 We set the Planck constant $\hbar=1$ throughout the paper. By $\hat{b}_{k}$ and $\hat{b}_{k}^{\dagger}$ we denoted, respectively, the annihilation and creation operators of the $k$-th  harmonic oscillator acting as
\begin{equation}
\hat{b}_{k}\ket{n}_{k}=\sqrt{n}\ket{n-1}_{k},
\end{equation}
\begin{equation}
\hat{b}_{k}^{\dagger}\ket{n}_{k}=\sqrt{(n+1)}\ket{n+1}_{k},
\end{equation}
where $\ket{n}_{k}$ is the number state vector in $\mathcal{H}_{\mathcal{E},k}$.
The operators $\hat{b}_{k}$ and $\hat{b}_{l}^{\dagger}$ satisfy the standard canonical commutation relations (CCR):
\begin{equation}\label{CCR1}
[\hat{b}_{k}, \hat{b}_{l}] = 0,\;\;\; [\hat{b}_{k}^{\dagger}, \hat{b}_{l}^{\dagger}] = 0,\;\;\;
[\hat{b}_{k}, \hat{b}_{l}^{\dagger}] =\delta_{kl}. 
\end{equation}
The model is formulated in the framework of standard assumptions made in quantum optics:  a flat coupling constant, rotating wave-approximation, and the extension of the lower limit of integration over frequency to minus infinity  \cite{Gross18,Ciccarello2017,Fischer2018a,Scully1997}. Thus we assume that the bandwidth of the spectrum is much smaller than the central frequency of the pulse. The Hamiltonian $\hat{H}_{k}$ is written in the interaction picture eliminating the free evolution of the field. By $\hat{H}_\mathcal{S}$ we denoted the Hamiltonian of the system $\mathcal{S}$. Further, $\hat{L}\in\mathcal{B}(\mathcal{H}_S)$ and $\hat{L}$ is called a jump or Lindblad operator. 
For  $\mathcal{S}$ being a two-level atom  $\hat{L}=\sqrt{\Gamma}\hat{\sigma}_{-}$, where $\Gamma$ is a positive coupling constant, and $\hat{\sigma}_{-}$ is a lowering operator of the atom. If $\mathcal{S}$ is a harmonic oscillator (a cavity mode) then  $\hat{L}=\sqrt{\Gamma}\hat{a}$, where $\hat{a}$ is the annihilation operator. We shall use for (\ref{Vk}) the Fock representation such that 
\begin{equation}\label{vil}
\exp\left(-i\tau \hat{H}_{k}\right)  = \sum_{mm^{\prime}} \ket{m}_{k} \bra{ m^{\prime}}_{k} \otimes\hat{\mathbbm{1}}_{\mathcal{E}}^{[k+1} \otimes \hat{V}_{mm^{\prime}},
\end{equation}
where $\hat{V}_{mm^{\prime}} \in\mathcal{B}(\mathcal{H}_S)$ and $m,m^{\prime}=0,1,2,\ldots$. 

The unitary operator describing the discrete time evolution of the composed system from zero time up to $\tau j $ for $j\geq 1$ is defined as 
\begin{equation}
\hat{U}_{j} = \hat{\mathbb{V}}_{j-1} \hat{\mathbb{V}}_{j-2} \ldots \hat{\mathbb{V}}_{0},\;\;\;\;\;\hat{U}_{0}=\hat{\mathbbm{1}}.
\end{equation}
Note that it acts non-trivially on the space $\mathcal{H}_{\mathcal{E}}^{j]}\otimes \mathcal{H_{S}}$ and as an identity operator on $\mathcal{H}_{\mathcal{E}}^{[j+1}$. After the $j$th first interactions the state of the composed system defined in $\mathcal{H}_{\mathcal{E}}\otimes \mathcal{H_{S}}$ reads as
\begin{equation}
\rho \rightarrow \hat{U}_{j}\rho \hat{U}_{j}^{\dagger}. 
\end{equation}
Taking the partial trace over $\mathcal{E}$, we obtain the reduced state of $\mathcal{S}$:
\begin{equation}
\sigma_{j}=\mathrm{Tr}_{\mathcal{E}}[\hat{U}_{j}\rho \hat{U}_{j}^{\dagger}].
\end{equation}

\section{Squeezed number states of the environment}

We start this section with the introduction of the creation wave-packet operator acting in the Hilbert space $\mathcal{H}_{\mathcal{E}}^{[j}$:
\begin{equation}\label{creation}
\hat{B}_{[j}^{\dagger}[\xi]=\sum_{k=j}^{M-1}\sqrt{\tau}\xi_{k}\hat{\tilde{b}}_{k}^{\dagger},
\end{equation}
where 
\begin{equation}
\hat{\tilde{b}}_{k}^{\dagger}=\hat{\mathbbm{1}}_\mathcal{E}^{k-1]} \otimes \hat{b}_{k}^{\dagger}\otimes\hat{\mathbbm{1}}_\mathcal{E}^{[k+1},
\end{equation}
${\xi}_{k}\in \mathbbm{C}$, and 
$\sum_{k=0}^{M-1} \tau\lvert {\xi}_{k}\rvert^2=1$. 
The commutator of  $\hat{B}_{[j}^{\dagger}[\xi]$ and its Hermitian-conjugate operator $\hat{B}_{[j}[\xi]$ is  obtained, with the help of CCR (\ref{CCR1}), as 
\begin{equation}
[\hat{B}_{[j}[\xi],\hat{B}_{[j}^{\dagger}[\xi]]=\sum_{k=j}^{M-1}\tau \lvert\xi_{k}\rvert^2.
\end{equation}
The creation operator (\ref{creation}) can be used to construct the number vectors
\begin{equation}\label{vectorm}
\ket{m_{\xi}}_{[j} = \frac{1}{\sqrt{m!}}\left(\hat{B}_{[j}^{\dagger}[\xi]\right)^{m}\ket{vac}_{[j},
\end{equation}
where $\ket{vac}_{[j} =\ket{0}_{j}\otimes\ket{0}_{j+1}\otimes\ldots\ket{0}_{M-1}$ is the vacuum vector in $\mathcal{H}_{\mathcal{E}}^{[j}$ and $m=0,1,\ldots $. Clearly, $\ket{0_{\xi}}_{[j}=\ket{vac}_{[j}$. Let us observe that the number vectors are mutually orthogonal:  
\begin{equation}
_{[j}\bra{m^{\prime}_{\xi}}\ket{m_{\xi}}_{[j}=\delta_{m^{\prime}m}\left(\sum_{k=j}^{M-1}\tau\lvert \xi_{k} \rvert^2\right)^m
\end{equation}
and they possess the additive decomposition property \cite{Dabrowska19}
\begin{eqnarray}\label{adp}
\ket{m_{\xi}}_{[j}=\sum_{m^{\prime}=0}^{m}
\sqrt{\binom{m}{m^{\prime}}}
(\sqrt{\tau}\xi_{j})^{m^{\prime}}\ket{m^{\prime}}_{j}\otimes \ket{(m-m^{\prime})_{\xi}}_{[j+1}. 
\end{eqnarray}
One can check that the action of the creation operator $\hat{B}_{[j}^{\dagger}[\xi]$ on the number vector gives
\begin{equation}\label{prop1}
\hat{B}_{[j}^{\dagger}[\xi]\ket{m_{\xi}}_{[j} =\sqrt{m+1} \ket{(m+1)_{\xi}}_{[j}
\end{equation}
and for the annihilation operators, one finds
\begin{equation}\label{prop2}
\hat{\mathbbm{1}}_\mathcal{E}^{[j,k-1]} \otimes\, \hat{b}_{k}\otimes\hat{\mathbbm{1}}_\mathcal{E}^{[k+1}\ket{m_{\xi}}_{[j} =\sqrt{m\tau}\xi_{k}\ket{(m-1)_{\xi}}_{[j},
\end{equation}
\begin{equation}\label{prop3}
\hat{B}_{[j}[\xi]\ket{m_{\xi}}_{[j} =\sqrt{m} \sum_{k=j}^{M-1}\tau\lvert\xi_{k}\rvert^2\ket{(m-1)_{\xi}}_{[j}. 
\end{equation}

In this paper we consider the number state defined by
\begin{equation}\label{nphot}
\ket{n_{\xi}} = \frac{1}{\sqrt{n!}}\left(\hat{B}^{\dagger}[\xi]\right)^{n}\ket{vac},
\end{equation}
where $\ket{vac}=\ket{0}_{0}\otimes\ket{0}_{1}\otimes\ldots\ket{0}_{M-1}$ is the vacuum vector in $\mathcal{H}_{\mathcal{E}}$,  and $\hat{B}^{\dagger}[\xi]=\hat{B}_{[0}^{\dagger}[\xi]$. Clearly, the number state is an eigenvector of the photon number operator, defined in $\mathcal{H}_{\mathcal{E}}$ by
\begin{equation}
\hat{n}=\sum_{k=0}^{M-1} \hat{\tilde{b}}_{k}^{\dagger}\hat{\tilde{b}}_{k}.
\end{equation}
Note that (\ref{nphot}) is an entangled state of the bath harmonic oscillators and a degree of this entanglement depends on $\xi$. Clearly, we deal here with $n$ independent photons of the same profiles $\xi$.
Let us notice that the state (\ref{nphot}) is a discrete analogue of the continuous-mode number state discussed, for instance, in \cite{BLPS90,Loudon2000,O06,RMS07}.

We define the squeezed number state in $\mathcal{H}_{\mathcal{E}}$ by the formula
\begin{equation}\label{nsqueezed}
\ket{n_{\gamma,\xi}} = \hat{S}[\gamma,\xi]\ket{n_{\xi}}
\end{equation}
with the squeeze operator
\begin{equation}\label{squeezing}
\hat{S}[\gamma,\xi]=\exp \left(\gamma\hat{B}^2[\xi]-\gamma^{\ast}\hat{B}^{\dagger 2}[\xi]\right),
\end{equation}
where $\gamma=\frac{r}{2}e^{-2i\phi}$. Of course, the squeezed number states are mutually orthogonal:
\begin{equation}
\bra{n^{\prime}_{\gamma,\xi}}\ket{n_{\gamma,\xi}^{\prime\prime}}=\delta_{n^{\prime}n^{\prime\prime}},
\end{equation} 
where $n^{\prime},n^{\prime\prime}\in \mathbb{N}$. Using the squeeze operator (\ref{squeezing}), we obtain the Bogoliubov transformation:
\begin{equation}\label{transformation}
\hat{S}^{\dagger}[\gamma,\xi]\;\hat{\tilde{b}}_{k}\hat{S}[\gamma,\xi]=\hat{\tilde{b}}_{k}
+\sqrt{\tau}\xi_{k}\left((c-1)\hat{B}[\xi]-se^{2i\phi}\hat{B}^{\dagger}[\xi]\right)
\end{equation}
where $c= \cosh r$ and $s=\sinh r$. Thus the unitary transformation of the wave-packet annihilation operator is given as
\begin{equation}\label{transformation2}
\hat{S}^{\dagger}[\gamma,\xi]\hat{B}[\xi]S[\gamma,\xi]= 
c\hat{B}[\xi]-
se^{2i\phi}\hat{B}^{\dagger}[\xi].
\end{equation}
By (\ref{transformation}) and the properties (\ref{prop1})-(\ref{prop3}) we get the mean value
\begin{equation}
\bra{ n_{\gamma,\xi}}\hat{\tilde{b}}_{k}\ket{n_{\gamma,\xi}}=0.
\end{equation}
Similar calculations allows us to find the two-time correlation functions for the field:
\begin{equation}
\bra{n_{\gamma,\xi}}\hat{\tilde{b}}^{\dagger}_{k}\hat{\tilde{b}}_{l}\ket{n_{\gamma,\xi}}=\tau \xi_{k}^{\ast}\xi_{l}\left(c^2n+s^2(n+1)\right).
\end{equation}
Let us notice that it is a function depending on the times $\tau k$ and $\tau l$. The mean number of photons in the field in the time interval $[k\tau,(k+1)\tau)$ is given by 
\begin{equation}
\bra{n_{\gamma,\xi}}\hat{\tilde{b}}^{\dagger}_{k}\hat{\tilde{b}}_{k}\ket{n_{\gamma,\xi}}=\tau \lvert\xi_{k}\rvert^2\left(c^2n+s^2(n+1)\right)
\end{equation}
and for the whole wave-packet prepared in the squeezed number state we have
\begin{equation}
\bra{ n_{\gamma,\xi}}\hat{n}\ket{n_{\gamma,\xi}}=c^2n+s^2(n+1).
\end{equation}
One can easily check that the expectation values of the quadrature operators,
\begin{equation}
\hat{X}[\xi]=\frac{1}{\sqrt{2}}\left(\hat{B}[\xi]e^{-i\phi}+\hat{B}[\xi]e^{i\phi}\right),
\end{equation}
\begin{equation}
\hat{Y}[\xi]=\frac{-i}{\sqrt{2}}\left(\hat{B}[\xi]e^{-i\phi}-\hat{B}[\xi]e^{i\phi}\right),
\end{equation}
are equaled to zero for the squeezed number state while their variances for $\phi=0$ have the form
\begin{equation}
\bra{n_{\gamma,\xi}}(\Delta \hat{X}[\xi])^2\ket{n_{\gamma,\xi}}=\left(2 n+1\right)e^{-2r}, 
\end{equation}
\begin{equation}
\bra{n_{\gamma,\xi}} (\Delta \hat{Y}[\xi])^2\ket{n_{\gamma,\xi}}=\left(2 n+1\right)e^{2r}.
\end{equation}

Note that any squeezed Fock state $\ket{n_{\gamma,\xi}}$ can be expanded into the number states \cite{Kim1989}:
\begin{equation}\label{ndecomposition}
\ket{n_{\gamma,\xi}}=\sum_{m=0}^{+\infty}a_{m}(n_\gamma)\ket{m_{\xi}}.
\end{equation} 
For instance, the squeezed vacuum state can be expressed as \cite{Kim1989}
\begin{equation}
\ket{0_{\gamma, \xi}}= \sum_{n=0}^{+\infty}\frac{(-1)^ne^{2i\phi n}}{\sqrt{n!\cosh r}}\left(\frac{\tanh r}{2}\right)^n\sqrt{(-1)^n H_{2n}(0)}\ket{(2n)_{\xi}},
\end{equation}    
where $H_{2n}(0)$ is the value for $x=0$ of the Hermite polynomial 
\begin{equation}
H_{2n}(x)=(2n)!\sum_{m=0}^{n}(-1)^{m}\frac{(2x)^{2n-2m}}{m!(2n-2m)!}. 
\end{equation}

The coefficients in the expansion of squeezed number states are connected by the following relation 
\begin{equation}\label{rel1}
\sqrt{m+1}a_{m+1}(n_\gamma)=\sqrt{n}c\;a_{m}((n-1)_\gamma)-\sqrt{n+1}se^{2i\phi}a_{m}((n+1)_\gamma).
\end{equation}
To show the above equation one needs to write down the action of wave-packet annihilation operator $\hat{B}[\xi]$ on the squeezed number state applying the transformation (\ref{transformation2}), then expand both sides of the equality in the basis of number states, and compare the coefficients for independent vectors on both sides.  

\begin{Remark}
	An arbitrary photon number state vector in $\mathcal{H_{E}}$ can be defined as 
	\begin{equation}\label{arbitrary}
	\ket{n_{\pmb{\varphi}}}=\frac{1}{\mathcal{N}}\sum_{k_{1},k_{2},\ldots,k_{n}=0}^{M-1}
	\tau^{n/2}\varphi_{k_{n}\ldots k_{2} k_{1}}\hat{\tilde{b}}_{k_{n}}^{\dagger}\ldots \hat{\tilde{b}}_{k_{2}}^{\dagger}\hat{\tilde{b}}_{k_{1}}^{\dagger}\ket{vac},
	\end{equation}
	where ${\mathcal{N}}$ is the normalization factor. 
	We do not assume any symmetry property for a  tensor $\varphi_{k_{n}\ldots k_{2} k_{1}}$.
	In particular situation
	\begin{equation}
	\varphi_{k_{n}\ldots k_{2} k_{1}} = \xi^{(n)}_{k_n} \ldots \xi^{(1)}_{k_1} ,
	\end{equation}
	where $\xi^{(i)}_{k_i}$ are the profiles of $n$ photons. In this paper we consider the case when all profiles are the same, so the photons are indistinguishable.
\end{Remark}

\begin{Remark}
In a general case, a squeeze operator in $\mathcal{H_{E}}$ can be defined by
\begin{equation}
\hat{S}[\pmb{\zeta}] = \exp\left(\hat{P}[\pmb{\zeta}]-\hat{P}^{\dagger}[\pmb{\zeta}]\right),
\end{equation}
where
\begin{equation}
\hat{P}[\pmb{\zeta}]=\sum_{k_1,k_2=0}^{M-1}\zeta_{k_2,k_1}\hat{\tilde{b}}_{k_2}\hat{\tilde{b}}_{k_1}.
\end{equation}
\end{Remark}

We assume that the initial state of the composed $\mathcal{E}+\mathcal{S}$ system has the form
\begin{equation}\label{ini0}
\ket{\Psi_{0}}=\ket{n_{\gamma,\xi}}\otimes\ket{\psi_{0}},
\end{equation}
where $\ket{\psi_{0}}$ is the initial state of $\mathcal{S}$.

\section{Conditioning in a repeated measurement model}

The environment $\mathcal{E}$ can be considered as a measuring device providing information on $\mathcal{S}$.  We describe in this Section the conditional state of the compound system depending on the results of the measurement performed on the elements of the output field. We assume that after each interaction a measurement is performed on the last harmonic oscillator which has interacted with the system $\mathcal{S}$. We consider the measurement of the field observable
\begin{equation}\label{obs}
\hat{n}_{k}=\hat{b}_{k}^{\dagger}\hat{b}_{k}=\ket{n}_{k} \bra{ n}_{k}, \;\;\; k=0,1,\ldots.
\end{equation} 
We assume that a detector is perfect and it works instantaneously. To represent the results of measurements performed up to time $j\tau$ we use the stochastic vector $\pmb{\eta}_j=(\eta_j,\eta_{j-1},\ldots,\eta_1)$.

\begin{theorem}\label{TH-1} The conditional ({\it a posteriori}) state vector of the system $\mathcal{S}$ and the input part of the environment (the part of the environment which has not interacted with $\mathcal{S}$ up to $j\tau$) for the initial state (\ref{ini0}) and the measurement of the observable (\ref{obs}) at time $j\tau$  is given by
\begin{equation}\label{condst1}
\ket*{\tilde{\Psi}_{j\vert \pmb{\eta}_j}^{n}} 
= \frac{\ket*{\Psi_{j\vert \pmb{ \eta}_j}^{n}}}
{\sqrt{\bra*{\Psi_{j\vert \pmb{\eta}_j}^{n}}\ket*{\Psi_{j\vert \pmb{\eta}_j}^{n}}}},
\end{equation}	
where $\ket*{\Psi_{j\vert \pmb{ \eta}_j}^{n}}$ is the unnormalized conditional vector from $\mathcal{H}_\mathcal{E}^{[j} \otimes \mathcal{H}_\mathcal{S}$ having the form
\begin{equation}\label{condst2}
\ket*{\Psi_{j\vert\pmb{\eta}_j}^{n}}=
\sum_{m=0}^{+\infty}\ket*{m_{\xi}}_{[j}\otimes\ket*{\psi_{j\vert\pmb{\eta}_j}^{n}(m)},
\end{equation}
where $\{\ket*{\psi_{j\vert\pmb{\eta}_j}^{n}(m)}\}$, $m=0,1,\ldots$ is the set of conditional vectors from $\mathcal{H}_{\mathcal{S}}$ which satisfy the set of recurrence equations
\begin{equation}\label{rec}
\ket*{\psi_{j+1\vert\pmb{\eta}_{j+1}}^{n}(m)} = \sum_{m^{\prime}=0}^{+\infty} \sqrt{\binom{m+m^{\prime}}{m^{\prime}}} \left(\sqrt{\tau} \xi_j\right)^{m^{\prime}} \hat{V}_{\eta_{j+1} m^{\prime}} \ket*{\psi_{j\vert\pmb{\eta}_j}^{n}(m+m^{\prime})},
\end{equation}
The operators $\hat{V}_{\eta_{j+1} r}\in\mathcal{B}(\mathcal{H}_S)$ are defined by (\ref{vil}), and initially we have
\begin{equation}
\ket*{\psi^{n}_{j=0}(m)}=a_{m}(n_\gamma)\ket{\psi_{0}}. 
\end{equation}
For the proof see \ref{B}.
\end{theorem}
Let us emphasize that the infinite set of conditional vectors $\{\ket*{\psi_{j\vert\pmb{\eta}_j}^{n}(m)}\}$ with $m=0,1,\ldots$ depends on the initial state of the composed system and all results of the measurements up to $j\tau$. It is seen from the structure of (\ref{condst2}) that the {\it a posteriori} state (\ref{condst1}) is an entangled state of the system $\mathcal{S}$ and the input part of environment. This entanglement makes the evolution of the system $\mathcal{S}$ non-Markovian.

The conditional probability of detecting $m$ photons at $(j+1)\tau$ given the results of all measurements up to $j\tau$ is defined by
\begin{equation}
p_{j+1}\left(m\vert \,\ket*{\tilde{\Psi}_{j\vert \pmb{\eta}_j}^{n}}\right)=\frac{\bra*{\Psi_{j\vert \pmb{\eta}_j}^{n}}  \hat{\mathbb{V}}_{[j}^{\dagger}\left(\ket{m}_{j}\bra{ m}_{j}\otimes \mathbbm{1}_\mathcal{E}^{[j+1}\otimes \mathbbm{1}_\mathcal{S}\right) \hat{\mathbb{V}}_{[j}  \ket*{\Psi_{j\vert \pmb{\eta}_j}^{n}}}{\bra*{\Psi_{j\vert \pmb{\eta}_j}^{n}}\ket*{\Psi_{j\vert \pmb{\eta}_j}^{n}}}.
\end{equation}
By the property $\hat{V}_{mr}=\hat{O}(\sqrt{\tau}^{\lvert m-r\rvert})$, we obtain for a detection of zero photons 
\begin{equation}
p_{j+1}\left(0\vert\,\ket*{\tilde{\Psi}_{j\vert \pmb{\eta}_j}^{n}}\right)=1+\overline{O}(\tau),
\end{equation}
and for $m\geq 1$ photons
\begin{equation}
p_{j+1}\left(m\vert\,\ket*{\tilde{\Psi}_{j\vert \pmb{\eta}_j}^{n}}\right)=\overline{O}(\tau^m),
\end{equation}
where $\overline{O}(.)$ is the Landau symbol. Therefore the probability of detecting more than one photon in the output field in the time interval $[k\tau, (k+1)\tau)$ is an expression of order $\overline{O}(\tau^2)$.  The probability of such detection is equal to zero in the continuous-time limit and we ignore the such cases. Now by neglecting in (\ref{rec}) all terms of order more than one in $\tau$  and the terms associated with the processes of probability of $\overline{O}(\tau^2)$, we obtain from (\ref{rec}) the following the set of difference equations 
\begin{eqnarray}\label{rec2}
\lefteqn{\ket*{\psi_{j+1\vert\pmb{\eta}_{j+1}}^{n}(m)}=}\\  &=& \hat{V}_{\eta_{j+1} 0} \ket*{\psi_{j\vert\pmb{\eta}_j}^{n}(m)}+\sqrt{(m+1)\tau} \xi_j \hat{V}_{\eta_{j+1} 1} \ket*{\psi_{j\vert\pmb{\eta}_j}^{n}(m+1)} \nonumber
\end{eqnarray}
with the system operators
\begin{eqnarray}\label{vmatrix}
\hat{V}_{00}&=& \hat{\mathbbm{1}}_{\mathcal{S}} - i\tau \hat{H}_{\mathcal{S}} -  \frac{\tau}{2}\hat{L}^\dagger \hat{L} + \hat{O}(\tau^{2}) ,\nonumber\\
\hat{V}_{10}&=&\sqrt{\tau} \hat{L} + \hat{O}(\tau^{3/2}) ,\nonumber\\
\hat{V}_{01}&=&- \sqrt{\tau} \hat{L}^\dagger + \hat{O}(\tau^{3/2}),\nonumber\\
\hat{V}_{11}&=& \hat{\mathbbm{1}}_{\mathcal{S}} + \hat{O}(\tau).
\end{eqnarray}
Note that here the random vector $\pmb{\eta}_j$ consists only of zeros and ones. Thus the system interacting with the field prepared in the state $\ket*{n_{\gamma,\xi}}$ can absorb or emit at most one photon in the interval of the length $\tau$. The processes of absorption and emission of more than one photon are not considered because their probabilities are of order $\overline{O}(\tau^2)$.

\section{Discrete and continuous evolution equations}

We shall show that a recurrence recipe for the state of $\mathcal{S}$ is given by an infinite set of coupled equations. 
A sequence of measurements performed on the output field gives rise to  the stochastic evolution of the system $\mathcal{S}$. In this Section, we 
describe conditional as well as unconditional dynamics of $\mathcal{S}$ in a discrete and continuous in-time cases. 

Let us start from establishing a set of stochastic discrete evolution equations for $\mathcal{S}$.   
To this aim, we take the partial trace of the random operator $\ket*{\tilde{\Psi}_{j\vert \pmb{\eta}_j}^{n}} \bra*{\tilde{\Psi}_{j\vert \pmb{\eta}_j}^{n}}$ over the environmental part. One can easily check that the reduced state of the system $\mathcal{S}$ at the time $j\tau$ has  the form 
\begin{equation}\label{condS}
\tilde{\rho}_{j\vert  \pmb{\eta}_{j}}
=\frac{\rho_{j\vert  \pmb{\eta}_{j}}}{\mathrm{Tr}_{\mathcal{S}}\rho_{j\vert  \pmb{\eta}_{j}}},
\end{equation}
where
\begin{equation}\label{condst3}
\rho_{j\vert  \pmb{\eta}_{j}}=\sum_{m=0}^{+\infty}\ket*{\psi_{j\vert  \pmb{\eta}_{j}}^{n}(m)}\bra*{\psi_{j\vert  \pmb{\eta}_{j}}^{n}(m)} \left(\sum_{k=j}^{N-1}\tau\lvert \xi_{k}\rvert^2\right)^m.
\end{equation}
Initially $\rho_{j=0}=\ket*{\psi_{0}}\bra{\psi_{0}}$. Note that the quantum repeated interactions and measurements provides a sequence of random states $\tilde{\rho}_{j\vert  \pmb{\eta}_{j}}$, $j=1,2,\ldots$ of the system $\mathcal{S}$ defining a discrete quantum trajectory on $\mathcal{H_{\mathcal{S}}}$. The operator $\tilde{\rho}_{j\vert  \pmb{\eta}_{j}}$ is the conditional state of $\mathcal{S}$ depending on the results of all measurements performed on the output field up to time $j\tau$, and the quantity 
\begin{equation}\label{probtra}
\mathrm{Tr}_{\mathcal{S}}\rho_{j\vert  \pmb{\eta}_{j}}
\end{equation} is the probability of a given trajectory. 

We shall derive a difference stochastic equation for $\tilde{\rho}_{j\vert  \pmb{\eta}_{j}}$. It is a recurrence recipe for the conditional state of $\mathcal{S}$ at the time $(j+1)\tau$ depending on the conditional state of $\mathcal{S}$ at $j\tau$ and a result of the measurement at $(j+1)\tau$. For this purpose we introduce operators
\begin{equation}\label{rhonnprime}
\rho_{j\vert\pmb{\eta}_j}^{n^{\prime},n^{\prime\prime}}=\sum_{m=0}^{+\infty}
\ket*{\psi_{j\vert  \pmb{\eta}_{j}}^{n^{\prime}}(m)}\bra*{\psi_{j\vert \pmb{\eta}_{j}}^{n^{\prime\prime}}(m)} \left(\sum_{k=j}^{M-1}\tau\lvert \xi_{k}\rvert^2\right)^m,
\end{equation}
where $n^{\prime}, n^{\prime\prime}\in \mathbb{N}$. Here $\{\ket*{\psi_{j\vert  \pmb{\eta}_{j}}^{n^{\prime}}(m)}\}$ for $m= 0,1,2,\ldots$ are the conditional vectors from $\mathcal{H_S}$, satisfying the recurrence equations of type of  (\ref{rec2}) and initially  
 $\ket*{\psi_{j=0}^{n^{\prime}}(m)}=a_{m}(n^{\prime}_\gamma)\ket*{\psi_{0}}$ with the coefficients $\{a_{m}(n^{\prime}_\gamma)\}$ specified by the expansion $\ket*{n_{\gamma,\xi}^{\prime}}=\sum_{m=0}^{+\infty}a_{m}(n_\gamma^{\prime})\ket*{m_{\xi}}$. Note that we consider different sets of conditional vectors related to different squeezed number states with a fixed parameter $\gamma$ and function $\xi$. One can check  that $\rho_{j\vert\pmb{\eta}_j}^{n^{\prime},n^{\prime\prime}}=\left(\rho_{j\vert\pmb{\eta}_j}^{n^{\prime\prime},n^{\prime}}\right)^{\dagger}$, $\rho_{j=0}^{n^{\prime},n^{\prime\prime}}=\delta_{n^{\prime}, n^{\prime\prime}}\ket*{\psi_{0}}\bra*{\psi_{0}}$, and $\rho_{j\vert\pmb{\eta}_j}^{n,n}=\rho_{j\vert\pmb{\eta}_j}$. To simplify our notation we drop the condition $\pmb{\eta}_j$ from now on.

We first determine the recurrence equations for the operators (\ref{rhonnprime}) for the case when the measurement gives us zero photons. If the result of measurement at time $(j+1)\tau$ is $\eta_{j+1}=0$, then referring to Eq. (\ref{rec2}) and the relation
\begin{eqnarray}\label{normm}
\|\ket*{m_{\xi}}_{[j+1}\|^2=\|\ket*{m_{\xi}}_{[j}\|^2-m\tau\lvert\xi_{j}\rvert^2\|\ket*{(m-1)_{\xi}}_{[j}\|^2+\overline{O}(\tau^2)
\end{eqnarray}
following from (\ref{prop1}), where $\|\cdot\|=\sqrt{\langle \cdot \rangle}$, we obtain for the conditional operator at $(j+1)\tau$ the formula
\begin{eqnarray}\label{filter1}
\rho_{j+1}^{n^{\prime},n^{\prime\prime}} &=& \rho_{j}^{n^{\prime},n^{\prime\prime}}-i\tau[\hat{H}_{\mathcal{S}},\rho_{j}^{n^{\prime},n^{\prime\prime}}]-\frac{\tau}{2}\left\{\hat{L}^{\dagger}\hat{L},\rho_{j}^{n^{\prime},n^{\prime\prime}}\right\}\nonumber\\
&&-\tau\xi_{j} \hat{L}^{\dagger}\sum_{m=0}^{+\infty}\sqrt{m+1}
\ket*{\psi_{j}^{n^{\prime}}(m+1)}
\bra*{\psi_{j}^{n^{\prime\prime}}(m)}\left(\sum_{k=j}^{M-1}\tau\lvert\xi_{k}\rvert^2\right)^m \nonumber\\
&&-\tau\xi_{j}^{\ast} \sum_{m=0}^{+\infty}\ket*{\psi_{j}^{n^{\prime}}(m)}\bra*{\psi_{j}^{n^{\prime\prime}}(m+1)}\sqrt{m+1}\left(\sum_{k=j}^{M-1}\tau\lvert \xi_{k}\rvert^2\right)^m \hat{L}\\
&&-\tau\lvert \xi_{j}\rvert^2 \sum_{m=0}^{+\infty}(m+1)\ket*{\psi_{j}^{n^{\prime}}(m+1)}\bra*{\psi_{j}^{n^{\prime\prime}}(m+1)}\left(\sum_{k=j}^{M-1}\tau\lvert\xi_{k}\rvert^2\right)^{m}+ \hat{O}(\tau^2)\nonumber
\end{eqnarray}
where $n^{\prime}, n^{\prime\prime}\in \mathbb{N}$. Now it is worth to observe that for any $n^{\prime}\in \mathbb{N}$ the conditional vectors satisfy the relation
\begin{equation}\label{rel}
\sqrt{m+1}\ket*{\psi_{j}^{n^{\prime}}(m+1)}=\sqrt{n^{\prime}}c\;\ket*{\psi_{j}^{n^{\prime}-1}(m)}-\sqrt{n^{\prime}+1}se^{2i\phi}\ket*{\psi_{j}^{n^{\prime}+1}(m)}. 
\end{equation}
To prove it one can use the general solution to the set of Eqs. (\ref{rec2}), given in the next Section, and the property (\ref{rel1}). With the help of (\ref{rel}) Eq. (\ref{filter1}) can be rewritten in the form
\begin{eqnarray}\label{filter2}
\rho_{j+1}^{n^{\prime},n^{\prime\prime}} &=& \rho_{j}^{n^{\prime},n^{\prime\prime}}-i\tau[\hat{H}_{\mathcal{S}},\rho_{j}(n^{\prime},n^{\prime\prime})]-\frac{\tau}{2}\left\{\hat{L}^{\dagger}\hat{L},\rho_{j}(n^{\prime},n^{\prime\prime})\right\}\nonumber\\
&&-\tau\xi_{j} \hat{L}^{\dagger}\left(\sqrt{n^{\prime}}c\rho_{j}^{n^{\prime}-1,n^{\prime\prime}}-\sqrt{n^{\prime}+1}se^{2i\phi}\rho_{j}^{n^{\prime}+1,n^{\prime\prime}}\right) \nonumber\\
&&-\tau\xi_{j}^{\ast}\left(\sqrt{n^{\prime\prime}}c\rho_{j}^{n^{\prime},n^{\prime\prime}-1}-\sqrt{n^{\prime\prime}+1}se^{-2i\phi}\rho_{j}^{n^{\prime},n^{\prime\prime}+1}\right) \hat{L} \nonumber\\
&&-\tau\lvert\xi_{j}\rvert^2\left(\sqrt{n^{\prime}n^{\prime\prime}}c^2\rho_{j}^{n^{\prime}-1,n^{\prime\prime}-1}+\sqrt{(n^{\prime}+1)(n^{\prime\prime}+1)}s^{2}\rho_{j}^{n^{\prime}+1,n^{\prime\prime}+1}\right)\nonumber\\
&&+\tau\lvert\xi_{j}\rvert^2 \sqrt{(n^{\prime}+1)n^{\prime\prime}}cse^{2i\phi}\rho_{j}^{n^{\prime}+1,n^{\prime\prime}-1}\nonumber\\
&&+\tau\lvert\xi_{j}\rvert^2 \sqrt{n^{\prime}(n^{\prime\prime}+1)}cse^{-2i\phi}\rho_{j}^{n^{\prime}-1,n^{\prime\prime}+1}+ \hat{O}(\tau^2). 
\end{eqnarray}

The conditional probability of detecting zero photons at the moment $(j+1)\tau$ given the {\it a posteriori} state of $\mathcal{S}$ at $j\tau$ was $\tilde{\rho}_{j}$ is defined as
\begin{equation}
p_{j+1}(0\vert\tilde{\rho}_{j})=\frac{\mathrm{Tr}_{\mathcal{S}}\rho_{j+1}}{\mathrm{Tr}_{\mathcal{S}}\rho_{j}}
\end{equation}
with $\rho_{j+1}$ stated by Eq. (\ref{filter2}).
One then finds that
\begin{equation}\label{prob0}
p_{j+1}(0\vert\tilde{\rho}_{j})=1- k_{j}\tau+\overline{O}(\tau^2),
\end{equation}
where
\begin{eqnarray}\label{intensity}
k_{j} &=& \mathrm{Tr}_{\mathcal{S}}\left\{ \hat{L}^{\dagger}\hat{L}\tilde{\rho}_{j}^{n,n}
+\xi_{j}\hat{L}^{\dagger}\left(\sqrt{n}c\tilde{\rho}_{j}^{n-1,n}-\sqrt{n+1}se^{2i\phi}\tilde{\rho}_{j}^{n+1,n}\right) \right. \nonumber\\
&&+\xi_{j}^{\ast}\hat{L}\left(\sqrt{n}c\tilde{\rho}_{j}^{n,n-1}-\sqrt{n+1}se^{-2i\phi}\tilde{\rho}_{j}^{n,n+1}\right)\nonumber\\ 
&&+\lvert\xi_{j}\rvert^2\left(nc^2\tilde{\rho}_{j}^{n-1,n-1}+(n+1)s^2\tilde{\rho}_{j}^{n+1,n+1}\right)\nonumber\\
&&\left.- \lvert\xi_{j}\rvert^2\sqrt{n(n+1)}cs\left(e^{2i\phi}\tilde{\rho}_{j}^{n+1,n-1}+e^{2i\phi}\tilde{\rho}_{j}^{n-1,n+1}\right)\right\}
\end{eqnarray}
and
\begin{equation}\label{oper}
\tilde{\rho}_{j}^{n^{\prime},n^{\prime\prime}}=\frac{\rho_{j}^{n^{\prime},n^{\prime\prime}}}{\mathrm{Tr}_{\mathcal{S}}\rho_{j}}.
\end{equation}
Clearly, $\tilde{\rho}_{j}^{n,n}$ is the conditional state of $\mathcal{S}$ at $j\tau$. Now, by 
\begin{equation}
\frac{1}{\mathrm{Tr}_{\mathcal{S}}\rho_{j+1}} =\frac{1}{ \mathrm{Tr}_{\mathcal{S}}\rho_j  }    \left(1+k_{j} \tau \right) + \overline{O}(\tau^2) ,
\end{equation}
we finally obtain from (\ref{filter2}) the infinite set of difference equations:
\begin{eqnarray}
\tilde{\rho}_{j+1}^{n^{\prime},n^{\prime\prime}} &=& \tilde{\rho_{j}}^{n^{\prime},n^{\prime\prime}}+\tau\left(k_{j}\tilde{\rho_{j}}^{n^{\prime},n^{\prime\prime}}-i[\hat{H}_{\mathcal{S}},\tilde{\rho}_{j}^{n^{\prime},n^{\prime\prime}}]-\frac{1}{2}\left\{\hat{L}^{\dagger}\hat{L},\tilde{\rho}_{j}^{n^{\prime},n^{\prime\prime}}\right\}\right)\nonumber\\
&&-\tau\xi_{j} \hat{L}^{\dagger}\left(\sqrt{n^{\prime}}c\tilde{\rho}_{j}^{n^{\prime}-1,n^{\prime\prime}}-\sqrt{n^{\prime}+1}se^{2i\phi}\tilde{\rho}_{j}^{n^{\prime}+1,n^{\prime\prime}}\right) \\
&&-\tau\xi_{j}^{\ast}\left(\sqrt{n^{\prime\prime}}c\tilde{\rho}_{j}^{n^{\prime},n^{\prime\prime}-1}-\sqrt{n^{\prime\prime}+1}se^{-2i\phi}\tilde{\rho}_{j}^{n^{\prime},n^{\prime\prime}+1}\right)  \hat{L} \nonumber\\
&&-\tau\lvert\xi_{j}\rvert^2 \left(\sqrt{n^{\prime}n^{\prime\prime}}c^2\tilde{\rho}_{j}^{n^{\prime}-1,n^{\prime\prime}-1}+\sqrt{(n^{\prime}+1)(n^{\prime\prime}+1)}s^{2}\tilde{\rho}_{j}^{n^{\prime}+1,n^{\prime\prime}+1}\right)\nonumber\\
&&+\tau\lvert\xi_{j}\rvert^2cs \left(\sqrt{(n^{\prime}+1)n^{\prime\prime}}e^{2i\phi}\tilde{\rho}_{j}^{n^{\prime}+1,n^{\prime\prime}-1}\right.\nonumber\\
&&\left.+\sqrt{n^{\prime}(n^{\prime\prime}+1)}e^{-2i\phi}\tilde{\rho}_{j}^{n^{\prime}-1,n^{\prime\prime}+1}\right)+ \hat{O}(\tau^2)\nonumber
\end{eqnarray}
for $n^{\prime},n^{\prime\prime}\in \mathbb{N}$.

Now we have to study the situation when at time $(j+1)\tau$ a detector register a photon. If $\eta_{j+1}=1$ then from Eqs. (\ref{rec2}) and (\ref{normm}) we obtain for the conditional operators the recurrence formula
\begin{eqnarray}\label{filter3}
\rho_{j+1}^{n^{\prime},n^{\prime\prime}} &=&\tau\left[\hat{L}\rho_{j}^{n^{\prime},n^{\prime\prime}}\hat{L}^{\dagger}+\xi_{j}\sum_{m=0}^{+\infty}\sqrt{m+1}\ket*{\psi^{n^{\prime}}_{j}(m+1)}\bra*{\psi^{n^{\prime\prime}}_{j}(m)}\sum_{k=j}^{M-1}\tau\lvert\xi_{k}\rvert^2\hat{L}^{\dagger}\right.\nonumber\\
&&+\xi_{j}^{\ast}\hat{L}\sum_{m=0}^{+\infty}\ket*{\psi^{n^{\prime}}_{j}(m)}\bra*{\psi^{n^{\prime\prime}}_{j}(m+1)}\sqrt{m+1}\sum_{k=j}^{M-1}\tau\lvert\xi_{k}\rvert^2\\
&&\left.+\lvert\xi_{j}\rvert^2\sum_{m=0}^{+\infty}(m+1)\ket*{\psi^{n^{\prime}}_{j}(m+1)}\bra*{\psi^{n^{\prime\prime}}_{j}(m+1)}\sum_{k=j}^{M-1}\tau\lvert\xi_{k}\rvert^2\right]+\hat{O}(\tau^2).\nonumber
\end{eqnarray}
Making use of (\ref{rel}) we obtain the equation
\begin{eqnarray}\label{filter4}
\rho_{j+1}^{n^{\prime},n^{\prime\prime}} &=& \tau\left[\hat{L}\rho_{j}^{n^{\prime},n^{\prime\prime}}\hat{L}^{\dagger}
+\xi_{j}^{\ast}\hat{L}\left(\sqrt{n^{\prime\prime}}c\rho_{j}^{n^{\prime},n^{\prime\prime}-1}
-\sqrt{n^{\prime\prime}+1}se^{-2i\phi}\rho_{j}^{n^{\prime},n^{\prime\prime}+1}\right)\right.\nonumber\\
&&+\xi_{j}\left(\sqrt{n^{\prime}}c\rho_{j}^{n^{\prime}-1,n^{\prime\prime}}
-\sqrt{n^{\prime}+1}se^{2i\phi}\rho_{j}^{n^{\prime}+1,n^{\prime\prime}}\right)
\hat{L}^{\dagger}\nonumber\\
&&+\lvert\xi_{j}\rvert^{2}\left(\sqrt{n^{\prime}n^{\prime\prime}}c^{2}\rho_{j}^{n^{\prime}-1,n^{\prime\prime}-1}
+\sqrt{(n^{\prime}+1)(n^{\prime\prime}+1)}s^{2}\rho_{j}^{n^{\prime}+1,n^{\prime\prime}+1}\right)\nonumber\\
&&-\lvert\xi_{j}\rvert^{2}cs
\left(\sqrt{n^{\prime}(n^{\prime\prime}+1)}e^{-2i\phi}\rho_{j}^{n^{\prime}-1,n^{\prime\prime}+1}
\right.\nonumber\\
&&\left.\left.+\sqrt{(n^{\prime}+1)n^{\prime\prime}}e^{2i\phi}\rho_{j}^{n^{\prime}+1,n^{\prime\prime}-1}\right)\right]+\hat{O}(\tau^2).
\end{eqnarray}

The conditional probability of detecting a photon at time $(j+1)\tau$ given the conditional state of $\mathcal{S}$ at $j\tau$ was $\tilde{\rho}_{j}$
is defined by
\begin{equation}
p_{j+1}(1\vert\tilde{\rho}_{j})=\frac{\mathrm{Tr}_{\mathcal{S}}\rho_{j+1}}{\mathrm{Tr}_{\mathcal{S}}\rho_{j}}
\end{equation}
with $\rho_{j+1}=\rho_{j+1}^{n,n}$ given by Eq. (\ref{filter3}), and one can find that
\begin{equation}\label{prob1}
p_{j+1}(1\vert\tilde{\rho}_{j})=k_{j}\tau.
\end{equation}
Hence for the case of detecting a photon at $(j+1)\tau$, we get for the difference equation of the form
\begin{eqnarray}\label{filter5}
\tilde{\rho}_{j+1}^{n^{\prime},n^{\prime\prime}} &=&k_{j}^{-1} \left[\hat{L}\tilde{\rho}_{j}^{n^{\prime},n^{\prime\prime}}\hat{L}^{\dagger}
+\xi_{j}^{\ast}\hat{L}\left(\sqrt{n^{\prime\prime}}c\tilde{\rho}_{j}^{n^{\prime},n^{\prime\prime}-1}
-\sqrt{n^{\prime\prime}+1}se^{-2i\phi}\tilde{\rho}_{j}^{n^{\prime},n^{\prime\prime}+1}\right)\right.\nonumber\\
&&+\xi_{j}\left(\sqrt{n^{\prime}}c\tilde{\rho}_{j}^{n^{\prime}-1,n^{\prime\prime}}
-\sqrt{n^{\prime}+1}se^{2i\phi}\tilde{\rho}_{j}^{n^{\prime}+1,n^{\prime\prime}}\right)
\hat{L}^{\dagger}\nonumber\\
&&+\lvert\xi_{j}\rvert^{2}\left(\sqrt{n^{\prime}n^{\prime\prime}}c^{2}\tilde{\rho}_{j}^{n^{\prime}-1,n^{\prime\prime}-1}
+\sqrt{(n^{\prime}+1)(n^{\prime\prime}+1)}s^{2}\tilde{\rho}_{j}^{n^{\prime}+1,n^{\prime\prime}+1}\right)\nonumber\\
&&-\lvert\xi_{j}\rvert^{2}cs
\left(\sqrt{n^{\prime}(n^{\prime\prime}+1)}e^{-2i\phi}\tilde{\rho}_{j}^{n^{\prime}-1,n^{\prime\prime}+1}\right.\nonumber\\&&
\left.\left.
+\sqrt{(n^{\prime}+1)n^{\prime\prime}}e^{2i\phi}\tilde{\rho}_{j}^{n^{\prime}+1,n^{\prime\prime}-1}\right)\right],
\end{eqnarray}
where $n^{\prime}, n^{\prime\prime}\in \mathbb{N}$. 

Let us introduce the stochastic discrete process 
\begin{equation}
N_{j}=\sum_{k=1}^{j}\eta_{k},
\end{equation}
with the increment defined as
\begin{equation}
\Delta N_{j}=N_{j+1}-N_{j}=\eta_{j+1}.
\end{equation}
One check, using (\ref{prob0}) and (\ref{prob1}), that the conditional mean value of $\Delta N_{j}$ has the form 
\begin{equation}\label{mean}
\mathbbm{E}[\Delta N_{j}\vert\tilde{\rho}_{j}]=k_{j}\tau+\overline{O}(\tau^2).
\end{equation}
Finally, by combining Eqs. (\ref{filter2}) and (\ref{filter5}) we obtain the set of filtering difference equations for the conditional operators $\tilde{\rho}_{j}^{n^{\prime},n^{\prime\prime}}$, where $n^{\prime}, n^{\prime\prime} \in \mathbb{N}$.

\begin{Proposition}\label{TH-2} 
The {\it a posteriori} evolution of the system $\mathcal{S}$ interacting with the environment prepared in the state (\ref{nsqueezed}) for the measurement of (\ref{obs}) is given by an infinite set of the coupled difference stochastic equations of the form
\begin{eqnarray}\label{filter6}
\tilde{\rho}_{j+1}^{n^{\prime},n^{\prime\prime}}&=& \tilde{\rho}_{j}^{n^{\prime},n^{\prime\prime}}+\mathcal{L}\tilde{\rho}_{j}^{n^{\prime},n^{\prime\prime}}\tau
+ [\sqrt{n^{\prime}}c\tilde{\rho}_{j}^{n^{\prime}-1,n^{\prime\prime}}
-\sqrt{n^{\prime}+1}se^{2i\phi}\tilde{\rho}_{j}^{n^{\prime}+1,n^{\prime\prime}},\hat{L}^{\dagger}]\xi_{j}\tau\nonumber\\
&&+[\hat{L}, \sqrt{n^{\prime\prime}}c\tilde{\rho}_{j}^{n^{\prime},n^{\prime\prime}-1}
-\sqrt{n^{\prime\prime}+1}se^{-2i\phi}\tilde{\rho}_{j}^{n^{\prime},n^{\prime\prime}+1}]\xi^{\ast}_{j}\tau \nonumber\\
&&+ \bigg\{\frac{1}{k_{j}}\left[\hat{L}\tilde{\rho}_{j}^{n^{\prime},n^{\prime\prime}}\hat{L}^{\dagger}
+\xi_{j}^{\ast}\hat{L}\left(\sqrt{n^{\prime\prime}}c\tilde{\rho}_{j}^{n^{\prime},n^{\prime\prime}-1}
-\sqrt{n^{\prime\prime}+1}se^{-2i\phi}\tilde{\rho}_{j}^{n^{\prime},n^{\prime\prime}+1}\right)\right.\nonumber\\
&&+\xi_{j}\left(\sqrt{n^{\prime}}c\tilde{\rho}_{j}^{n^{\prime}-1,n^{\prime\prime}}
-\sqrt{n^{\prime}+1}se^{2i\phi}\tilde{\rho}_{j}^{n^{\prime}+1,n^{\prime\prime}}\right)
\hat{L}^{\dagger}\nonumber\\
&&
+\lvert\xi_{j}\rvert^{2}\left(\sqrt{n^{\prime}n^{\prime\prime}}c^{2}\tilde{\rho}_{j}^{n^{\prime}-1,n^{\prime\prime}-1}
+\sqrt{(n^{\prime}+1)(n^{\prime\prime}+1)}s^{2}\tilde{\rho}_{j}^{n^{\prime}+1,n^{\prime\prime}+1}\right)\nonumber\\
&&-\lvert\xi_{j}\rvert^{2}cs
\sqrt{n^{\prime}(n^{\prime\prime}+1)}e^{-2i\phi}\tilde{\rho}_{j}^{n^{\prime}-1,n^{\prime\prime}+1}\nonumber\\
&&\left.-\lvert\xi_{j}\rvert^{2}cs\sqrt{(n^{\prime}+1)n^{\prime\prime}}e^{2i\phi}\tilde{\rho}_{j}^{n^{\prime}+1,n^{\prime\prime}-1}\right]-\tilde{\rho}_{j}^{n^{\prime},n^{\prime\prime}}\bigg\}\left(\Delta N_{j}-k_{j}\tau\right),
\end{eqnarray}
where  
\begin{equation}\label{supero}
\mathcal{L}\tilde{\rho}=-i[\hat{H}_{\mathcal{S}},\tilde{\rho}]-\frac{1}{2}\left\{\hat{L}^{\dagger}\hat{L},\tilde{\rho}\right\}
+\hat{L}\tilde{\rho} \hat{L}^{\dagger}
\end{equation}
 and the initial conditions: $\tilde{\rho}_{j=0}^{n^{\prime},n^{\prime\prime}}=\delta_{n^{\prime}, n^{\prime\prime}}\ket*{\psi_{0}}\bra*{\psi_{0}}$ for $n^{\prime}, n^{\prime\prime} \in \mathbb{N}$. Here $\{\hat{a},\hat{b}\}=\hat{a}\hat{b}+\hat{b}\hat{a}$. 
 The {\it a posteriori} state of $\mathcal{S}$ at time $j\tau$ is given by $\tilde{\rho}_{j}^{n,n}$. The discrete stochastic process $N_{j}$ characterizing photon counts defined by the observable (\ref{obs}) is uniquely determined by (\ref{prob0}) and (\ref{prob1})
\end{Proposition}
The filtring equations (\ref{filter6}) are nonlinear. If a photon was not registered at $(j+1)\tau$ then $\Delta N_{j}=0$ and (\ref{filter6}) reduces to (\ref{filter2}) and if a photon was detected at $(j+1)\tau$ then $\Delta N_{j}=1$, all terms proportional to $\tau$ in (\ref{filter6}) are negligible and (\ref{filter6}) is equivalent to (\ref{filter5}). We omitted in Eq. (\ref{filter6}) all terms that do not give contribution to the continuous situation when $\tau \to dt$.

Now taking the average over all trajectories we obtain from (\ref{filter6}) the unconditional ({\it a priori}) dynamics of the system $\mathcal{S}$. The {\it a priori} state of $\mathcal{S}$ at time $j\tau$ for $j>0$ can be obtained by taking the mean value of $\tilde{\rho}_{j}$ with respect to the measure defined by (\ref{probtra}):
\begin{equation}
\sigma_{j}^{n,n}=\langle \tilde{\rho}_{j}^{n,n}\rangle_{st}. 
\end{equation}
We take the stochastic mean of (\ref{filter6}) in the two steps. First we use the conditional mean of the increment $\Delta N_{j}$ and then we take the mean value of all the other elements on the past.
\begin{Proposition}
	The {\it a priori} dynamics of $\mathcal{S}$ is given by the infinite set of difference master equations
	\begin{eqnarray}\label{master1}
	{\sigma}_{j+1}^{n^{\prime},n^{\prime\prime}}&=& {\sigma}_{j}^{n^{\prime},n^{\prime\prime}}+\mathcal{L}\sigma_{j}^{n^{\prime},n^{\prime\prime}}\tau+ [\sqrt{n^{\prime}}c{\sigma}_{j}^{n^{\prime}-1,n^{\prime\prime}}
	-\sqrt{n^{\prime}+1}se^{2i\phi}{\sigma}_{j}^{n^{\prime}+1,n^{\prime\prime}},\hat{L}^{\dagger}]\xi_{j}\tau\nonumber\\
	&+&[\hat{L}, \sqrt{n^{\prime\prime}}c{\sigma}_{j}^{n^{\prime},n^{\prime\prime}-1}
	-\sqrt{n^{\prime\prime}+1}se^{-2i\phi}{\sigma}_{j}^{n^{\prime},n^{\prime\prime}+1}]\xi^{\ast}_{j}\tau 
	\end{eqnarray}
	where 
	\begin{equation}
	\sigma_{j}^{n^{\prime},n^{\prime\prime}}=\langle \tilde{\rho}_{j}^{n^{\prime},n^{\prime\prime}}\rangle_{st}
	\end{equation}
	with the initial condition ${\sigma}_{j=0}^{n^{\prime},n^{\prime\prime}}=\delta_{n^{\prime},n^{\prime\prime}}\ket*{\psi_{0}}\bra*{\psi_{0}}$ for $n^{\prime},n^{\prime\prime}\in \mathbb{N}$. 
\end{Proposition}

We end up this Section with the results for the continuous in-time dynamics. 
We take the limit of $\tau\to 0$ and $M\to \infty$ such that $T=M\tau$ is fixed. In this case we obtain from (\ref{filter6}) the infinite set of the coupled differential stochastic equations of the form
\begin{eqnarray}\label{filter7}
d\tilde{\rho}_{t}^{n^{\prime},n^{\prime\prime}}&=& \mathcal{L}\tilde{\rho}_{t}^{n^{\prime},n^{\prime\prime}}dt+ [\sqrt{n^{\prime}}c\tilde{\rho}_{t}^{n^{\prime}-1,n^{\prime\prime}}
-\sqrt{n^{\prime}+1}se^{2i\phi}\tilde{\rho}_{t}^{n^{\prime}+1,n^{\prime\prime}},\hat{L}^{\dagger}]\xi_{t}\tau\nonumber\\
&&+[\hat{L}, \sqrt{n^{\prime\prime}}c\tilde{\rho}_{t}^{n^{\prime},n^{\prime\prime}-1}
-\sqrt{n^{\prime\prime}+1}se^{-2i\phi}\tilde{\rho}_{t}^{n^{\prime},n^{\prime\prime}+1}]\xi^{\ast}_{t}\tau \nonumber\\
&&+ \bigg\{\frac{1}{k_{t}}\left[\hat{L}\tilde{\rho}_{t}^{n^{\prime},n^{\prime\prime}}\hat{L}^{\dagger}
+\xi_{t}^{\ast}\hat{L}\left(\sqrt{n^{\prime\prime}}c\tilde{\rho}_{t}^{n^{\prime},n^{\prime\prime}-1}
-\sqrt{n^{\prime\prime}+1}se^{-2i\phi}\tilde{\rho}_{t}^{n^{\prime},n^{\prime\prime}+1}\right)\right.\nonumber\\
&&+\xi_{t}\left(\sqrt{n^{\prime}}c\tilde{\rho}_{t}^{n^{\prime}-1,n^{\prime\prime}}
-\sqrt{n^{\prime}+1}se^{2i\phi}\tilde{\rho}_{t}^{n^{\prime}+1,n^{\prime\prime}}\right)
\hat{L}^{\dagger}\nonumber\\
&&+\lvert\xi_{t}\rvert^{2}\left(\sqrt{n^{\prime}n^{\prime\prime}}c^{2}\tilde{\rho}_{t}^{n^{\prime}-1,n^{\prime\prime}-1}
+\sqrt{(n^{\prime}+1)(n^{\prime\prime}+1)}s^{2}\tilde{\rho}_{t}^{n^{\prime}+1,n^{\prime\prime}+1}\right)\nonumber\\
&&-\lvert\xi_{t}\rvert^{2}cs
\sqrt{n^{\prime}(n^{\prime\prime}+1)}e^{-2i\phi}\tilde{\rho}_{t}^{n^{\prime}-1,n^{\prime\prime}+1}\nonumber\\
&&\left.-\lvert\xi_{t}\rvert^{2}cs\sqrt{(n^{\prime}+1)n^{\prime\prime}}e^{2i\phi}\tilde{\rho}_{t}^{n^{\prime}+1,n^{\prime\prime}-1}\right]-\tilde{\rho}_{t}^{n^{\prime},n^{\prime\prime}}\bigg\}\left(dN_{t}-k_{t}dt\right)
\end{eqnarray}
with 
\begin{eqnarray}\label{intensity2}
k_{t} &=& \mathrm{Tr}_{\mathcal{S}}\left\{ \hat{L}^{\dagger}\hat{L}\tilde{\rho}_{t}^{n,n}
+ \xi_{t}^{\ast}\hat{L}\left(\sqrt{n}c\tilde{\rho}_{t}^{n,n-1}-\sqrt{n+1}se^{-2i\phi}\tilde{\rho}_{t}^{n,n+1}\right)\right.\nonumber\\ 
&&+\xi_{t}\hat{L}^{\dagger}\left(\sqrt{n}c\tilde{\rho}_{t}^{n-1,n}-\sqrt{n+1}se^{2i\phi}\tilde{\rho}_{t}^{n+1,n}\right)  \nonumber\\
&&+\lvert\xi_{j}\rvert^2\left(nc^2\tilde{\rho}_{t}^{n-1,n-1}+(n+1)s^2\tilde{\rho}_{t}^{n+1,n+1}\right)\nonumber\\
&&\left.- \lvert\xi_{t}\rvert^2\sqrt{n(n+1)}cs\left(e^{2i\phi}\tilde{\rho}_{t}^{n+1,n-1}+e^{2i\phi}\tilde{\rho}_{t}^{n-1,n+1}\right)\right\}
\end{eqnarray}
and the initial condition of the form
$\tilde{\rho}_{t=0}^{n^{\prime},n^{\prime\prime}}=\delta_{n^{\prime}, n^{\prime\prime}}\ket*{\psi_{0}}\bra*{\psi_{0}}$. Here $N_{t}$ is the stochastic counting process with the increment $dN_{t}=N_{t+dt}-N_{t}$ having the conditional mean value
\begin{equation}\label{condmean}
\mathbbm{E}[dN_{t}\vert\tilde{\rho}_{t}]=k_{t}dt.
\end{equation}
For the process $N_{t}$ we get the relation $(dN_{t})^2=dN_{t}$, reflecting the fact that we can detect at most one photon in time interval $[t,t+dt)$. Note that $k_{t}$ is the intensity of $N_{t}$ and $k_{t}dt$ is the conditional mean number of photons detected from $t$ to $t+dt$. If we put $s=0$ and $c=1$, which means that we take $\gamma=0$, then (\ref{filter7}) reduce to the set of stochastic equations for the number photon state derived in \cite{Baragiola17,Dabrowska19}.  We would like to emphasise that the set of filtering equations (\ref{filter7}) agrees with the results published in \cite{Gross2022}.

Clearly, taking finally the limit of $T\to+\infty$, we get for $\xi\in L^{2}[0,\infty)$ the normalization condition
\begin{equation}
\int_{0}^{\infty}\lvert\xi_{t}\rvert^2dt=1.
\end{equation}

By taking the mean of (\ref{filter7}) we obtain the set of master differential equations describing the {\it a priori} dynamics of $\mathcal{S}$.  The reduced dynamics of $\mathcal{S}$ is given by the infinite set of differential equations
\begin{eqnarray}\label{master2}
\frac{d}{dt}{\sigma}_{t}^{n^{\prime},n^{\prime\prime}}&=&\mathcal{L}\sigma_{t}^{n^{\prime},n^{\prime\prime}}+ [\sqrt{n^{\prime}}c{\sigma}_{t}^{n^{\prime}-1,n^{\prime\prime}}
-\sqrt{n^{\prime}+1}se^{2i\phi}{\sigma}_{t}^{n^{\prime}+1,n^{\prime\prime}},\hat{L}^{\dagger}]\xi_{t}\nonumber\\
&+&[\hat{L}, \sqrt{n^{\prime\prime}}c{\sigma}_{t}^{n^{\prime},n^{\prime\prime}-1}
-\sqrt{n^{\prime\prime}+1}se^{-2i\phi}{\sigma}_{t}^{n^{\prime},n^{\prime\prime}+1}]\xi^{\ast}_{t}  
\end{eqnarray}
with the initial condition $\sigma_{t=0}^{n^{\prime},n^{\prime\prime}}=\delta_{n^{\prime}, n^{\prime\prime}}\ket*{\psi_{0}}\bra*{ \psi_{0}}$. The {\it a priori} state of $\mathcal{S}$ is given by $\sigma_{t}={\sigma}_{t}^{n,n}$. 	
It is clear that for the parameters $s=0$ and $c=1$ the infinite set of master equations (\ref{master2}) gives the set of $(n+1)^2$ equations for the number photon state \cite{Baragiola12,Baragiola17,Dabrowska19}. The set of master equations (\ref{master2}) one can derive directly from the set of master equations for the input field prepared in the number photon state, see \ref{C}.

\section{Quantum trajectories and photon statistics}

In this Section we present the solution to the set of stochastic master equations for discrete as well as continuous in-time dynamics. We display moreover the {\it a priori} state of the system by means of the quantum trajectories. The starting point in our discussion is the solution to the set of discrete stochastic equations (\ref{rec2}). 
We return in this Section to the notation with conditional subscripts.  

\subsection{Discrete case}

The general solution to the set of Eqs. (\ref{rec2}) can be written in the form
\begin{eqnarray}\label{sol}
\ket*{\psi^{n}_{j\vert\pmb{\eta}_{j}}(m)} &=& \left[\hat{V}_{\eta_{j} 0} \hat{V}_{\eta_{j-1} 0} \dots \hat{V}_{\eta_1 0}  a_{m}(n_{\gamma})\right.\\
&&\left.+\sum_{k=1}^{j}\sqrt{\frac{(m+k)!}{m!}}
\sum_{\pmb{ r} \in \mathbb{N}^j : \sum_{l} r_{l} = k}
\prod_{l=0}^{\stackrel{\longleftarrow}{j-1}}\sqrt{\tau}^{r_{l}} \xi_{l}^{r_{l}} \hat{V}_{\pmb{\eta}_{l+1}r_{l}} a_{m+k}(n_{\gamma})\right.]\ket*{\psi_{0}},\nonumber
\end{eqnarray}
where $m = 0, 1, 2, \ldots$, the vector $\pmb{r}$ consists only zeros and ones, and the arrow denotes time-ordered products. Let us stress that we get an infinite set of conditional vectors which depend on the initial state of the composed system and on all results of the measurement performed on the output field up to time $j\tau$. One can write down it for some explicit scenarios of detection. Instead of writing the whole sequence of all results from $0$ to $j\tau$, we indicate only moments of counts in the realization of stochastic vector $\pmb{\eta}_{j}=(\eta_j,\ldots,\eta_1)$. Thus the string $(l_s,\ldots,l_1)$, where $0<l_1<\ldots <l_s\leq j$, means that we detected exactly $s$ photons at moments $\tau_i = l_i\tau $ $(i=1,\ldots,s)$ and no other photons from time $0$ to $j\tau$. Of course, any $l_{i}\geq 1$. Let us introduce the operators:
\begin{equation}
\hat{A}_{i}=\hat{V}_{00}^{-i-1}\sqrt{\tau}\xi_{i}\hat{V}_{01}\hat{V}_{00}^{i},
\end{equation}
\begin{equation}
\hat{E}_{l}= \hat{V}_{00}^{-l}\hat{V}_{10}\hat{V}_{00}^{l-1},
\end{equation}
\begin{equation}
\hat{D}_{l}= \hat{V}_{00}^{-l}\sqrt{\tau}\xi_{l-1}\hat{V}_{11}\hat{V}_{00}^{l-1},
\end{equation}
where $i=0,1,2,\ldots$ and $l=1,2,\ldots$. Note that   
$\hat{A}_{i}$ is associated with an absorption of a photon by $\mathcal{S}$ at time $i\tau$, $\hat{E}_{l}$ with an emission of a photon by $\mathcal{S}$ at time $l\tau$, and $\hat{D}_{l}$ with a detection of a photon coming directly from the input field at time $l\tau$. Using these operators we can express the conditional vectors at time $j\tau$ for $m  = 0, 1, 2, \ldots$ as follows:
\begin{enumerate}
\item for detecting no photons from $0$ to $j\tau$, we obtain
\begin{eqnarray}\label{sol1}
\ket*{\psi^{n}_{j\vert0}(m)}&=& \hat{V}_{00}^{j}
\left[a_{m}(n_{\gamma})+\sum_{k=1}^{j}
\sqrt{\frac{(m+k)!}{m!}}\sum_{i_{k}=j-1}^{j-1}\sum_{i_{k-1}=j-2}^{i_{k}-1}\right.\nonumber\\
&&\left.\ldots\sum_{i_{1}=0}^{i_{2}-1}
\hat{A}_{i_{k}}\hat{A}_{i_{k-1}}\ldots \hat{A}_{i_{1}} a_{m+k}(n_{\gamma})\right]\ket*{\psi_{0}},
\end{eqnarray}
\item for a detection of a photon at time $l\tau$ and no other photons from $0$ to $j\tau$ we get:
\begin{equation}\label{sol2}
\ket*{\psi_{j\vert l}^{n}(m)}=\sum_{k=0}^{j}\sqrt{\frac{(m+k)!}{m!}}\hat{R}_{j\vert l}(k)a_{m+k}(n_{\gamma})\ket*{\psi_{0}},
\end{equation}
where
\begin{eqnarray}\label{sol3}
\hat{R}_{j\vert l}(0)&=& \hat{V}_{00}^{j}\hat{E}_{l},
\end{eqnarray}
\begin{eqnarray}\label{sol4}
\hat{R}_{j\vert l}(1)&=& \hat{V}_{00}^{j}\left(\hat{D}_{l}+
\hat{E}_{l}\sum_{i=0}^{l-2}\hat{A}_{i}+ \sum_{i=l}^{j-1}\hat{A}_{i}\hat{E}_{l}
\right)
\end{eqnarray}
\begin{eqnarray}\label{sol5}
\hat{R}_{j\vert l}(2)&=&\hat{V}_{00}^{j}\left(
\hat{D}_{l}\sum_{i=0}^{l-2}\hat{A}_{i}+\sum_{i=l}^{j-1}\hat{A}_{i}\hat{D}_{l}+
\hat{E}_{l}
\sum_{i_{2}=1}^{l-2}\sum_{i_{1}=0}^{i_{2}-1}\hat{A}_{i_2}\hat{A}_{i_1}\right.\nonumber\\
&+&\left.\sum_{i_{2}=l}^{j-1}\hat{A}_{i_2}\hat{E}_{l}
\sum_{i_{1}=0}^{l-2}\hat{A}_{i_1}+\sum_{i_{2}=l+1}^{j-1}\sum_{i_{1}=l}^{i_{2}-1}\hat{A}_{i_2}
\hat{A}_{i_1}\hat{E}_{l}
\right),
\end{eqnarray}
\begin{eqnarray}\label{sol6}
\hat{R}_{j\vert l}(3)&=&\hat{V}_{00}^{j}
\left(\hat{D}_{l}\sum_{i_{2}=1}^{l-2}\sum_{i_{1}=0}^{i_{2}-1}\hat{A}_{i_2}\hat{A}_{i_1}+
\sum_{i_{2}=l}^{j-1}\hat{A}_{i_2}\hat{D}_{l}\sum_{i_{1}=0}^{l-2}\hat{A}_{i_1}
\right.\nonumber\\
&&+\sum_{i_{2}=l+1}^{j-1}\sum_{i_{1}=l}^{i_{2}-1}\hat{A}_{i_2}\hat{A}_{i_1}\hat{D}_{l}+
\hat{E}_{l}\sum_{i_{3}=3}^{l-2}\sum_{i_{2}=2}^{i_3-1}\sum_{i_{1}=0}^{i_{2}-1}\hat{A}_{i_3}\hat{A}_{i_2}\hat{A}_{i_1}\nonumber\\
&&+\sum_{i_{3}=l}^{j-1}\hat{A}_{i_3}\hat{E}_{l}\sum_{i_{2}=1}^{l-2}\sum_{i_{1}=0}^{i_{2}-1}\hat{A}_{i_2}\hat{A}_{i_1}+\sum_{i_{2}=l+1}^{j-1}\sum_{i_{1}=l}^{i_{2}-1}\hat{A}_{i_2}\hat{A}_{i_1}\hat{E}_{l}\sum_{i=0}^{l-2}\hat{A}_{i}
\nonumber\\
&&\left.
+\sum_{i_{3}=l+3}^{j-1}\sum_{i_{2}=l+2}^{i_3-1}\sum_{i_{1}=l}^{i_{2}-1}\hat{A}_{i_3}\hat{A}_{i_2}\hat{A}_{i_1}\hat{E}_{l}
\right),
\end{eqnarray}
\end{enumerate}
and so on. In a general case, for detection of $s$ photons at times $l_1 \tau, l_2\tau, \ldots, l_s\tau$, where $l_1<l_2<\ldots<l_s$, and no other photons from $0$ to $j\tau$, the conditional vectors can be expressed as
\begin{equation}
\ket*{\psi_{j\vert l_s,\ldots, l_2,l_1}^{n}(m)}=\sum_{k=0}^{j}\sqrt{\frac{(m+k)!}{m!}}\hat{R}_{j\vert l_s,\ldots, l_2,l_1}(k)a_{m+k}(n_{\gamma})\ket*{\psi_{0}}
\end{equation}
The formula for the conditional operator $\hat{R}_{j\vert l_s,\ldots, l_2,l_1}(k)$ becomes more and more complicated with increasing numbers $k$ and $s$, but all expressions in this formula follows two simple rules:
\begin{enumerate}
	\item the number of photons emitted by $\mathcal{S}$ plus the number of detected photons which came directly from the input field is equaled to $s$:
	\begin{equation}
	n_E+n_D=s,
	\end{equation}
	\item the number of photons absorbed by $\mathcal{S}$ plus the number of detected photons which came directly from the input field is equaled to $k$:
	\begin{equation}
	n_A+n_D=k.
	\end{equation}
\end{enumerate}
The physical interpretations of the terms in the formula for $\hat{R}_{j\vert l_s,\ldots, l_2,l_1}(k)$ are not so complex. One can easily recognize there two sources of the detected photons: the input field and system $\mathcal{S}$. These two kinds of photons are indistinguishable for the observer. Clearly, for a particular system $\mathcal{S}$ not all terms give non-zero contribution to the conditional vector $\ket*{\psi_{j\vert l_s,\ldots, l_2,l_1}^{n}(m)}$. For instance, if $\mathcal{S}$ is a two-level atom all terms with the two or more successive absorptions disappear.

\subsection{Continuous case}

We would like to provide analytical formulae for the conditional as well as unconditional state of $\mathcal{S}$ and characterize the statistics of the output photons by means of the exclusive probability densities \cite{Srinivas81}. Let us notice that all realization of the counting stochastic process $N_{t}$ may be divided into disjoint sectors: $\mathcal{C}_s$ containing trajectories with exactly $s$ detected photons in the nonoverlapping intervals $[t_{1},t_{1}+dt_{1})$, $[t_{2},t_{2}+dt_{2})$, $\ldots$, $[t_{s},t_{s}+dt_{s})$ lying in the interval $(0,t]$, where $t_1 < t_{2}<\ldots < t_s$. In order to write down the conditional vectors in compact forms, we introduce the operators 
\begin{equation}\label{}
\hat{T}_t = e^{-i\hat{G}t},
\end{equation}
where $\hat{G}=\hat{H}_{\mathcal{S}}-\frac{i}{2}\hat{L}^{\dagger}\hat{L}$ is a non-Hermitian Hamiltonian,
\begin{equation}
\hat{A}_{t}=-\hat{T}_{-t}\xi_{t}\hat{L}^{\dagger}\hat{T}_{t},
\end{equation}
\begin{equation}\label{}
\hat{E}_t = \hat{T}_{-t}\hat{L}\hat{T}_{t}.
\end{equation}
Let us notice that $\hat{T}_t$ describes free propagation of $\mathcal{S}$ up to time $t$, $\hat{A}_{t}$ corresponds to an absorption of a photon by  $\mathcal{S}$ at $t$, and $\hat{E}_t$ to an emission of a photon by $\mathcal{S}$ at time $t$. 

In the continuous time limit we obtain the conditional vectors at time $t$ of the form:
\begin{enumerate}
\item from (\ref{sol1}) for zero photons from $0$ to $t$: 
	\begin{eqnarray}
	\ket*{\psi^{n}_{t\vert 0}(m)}&=&\hat{T}_{t}\left[a_{m}(n_{\gamma})+
	\sum_{k=1}^{+\infty}\sqrt{\frac{(m+k)!}{m!}}
	\int_{0}^{t}dt_{k}\int_{0}^{t_k}dt_{k-1} \right.\\&&\left.
	\ldots\int_{0}^{t_{3}}dt_{2}\int_{0}^{t_{2}}
	dt_{1} \hat{A}_{t_{k}}\hat{A}_{t_{k-1}}\ldots \hat{A}_{t_{2}}\hat{A}_{t_{1}}a_{m+k}(n_{\gamma})\right]\ket*{\psi_{0}},\nonumber
	\end{eqnarray}
	\item from (\ref{sol2}) for one photon detected at the time $t_1$ and no other counts in the interval $(0,t]$:	
\begin{equation}
\ket*{\psi_{t\vert t_{1}}^{n}(m)}=\sum_{k=0}^{+\infty}\sqrt{\frac{(m+k)!}{m!}}\hat{R}_{t\vert t_{1}}(k)a_{m+k}(n_{\gamma})\ket*{\psi_{0}},
\end{equation}
where from the expressions (\ref{sol3})-(\ref{sol5}), respectively, we get
\begin{equation}
\hat{R}_{t\vert t_{1}}(0)= \sqrt{dt_1}\hat{T}_{t} \,  \hat{E}_{t_1},
\end{equation}
\begin{eqnarray}
\hat{R}_{t\vert t_{1}}(1)&=& \sqrt{dt_1}\,  \hat{T}_t \Big[ \xi_{t_1} +  \hat{E}_{t_1} \int_0^{t_1} ds \hat{A}_s + \int_{t_1}^{t} ds \hat{A}_s \hat{E}_{t_1} \Big],
\end{eqnarray}
\begin{eqnarray}\label{veccon2}
R_{t\vert t_{1}}(2)&=& \sqrt{dt_1}\, \hat{T}_t \, \Big[ \xi_{t_1} \int_{0}^{t_1}ds \hat{A}_{s} +  \int_{t_1}^{t}ds \hat{A}_{s}\, \xi_{t_1} \nonumber\\
&&+\hat{E}_{t_1} \int_{0}^{t_1}ds^{\prime}\int_{0}^{s^{\prime}}ds \hat{A}_{s^{\prime}} \hat{A}_{s}+ \int_{t_1}^{t}ds^{\prime} \hat{A}_{s^{\prime}} \, \hat{E}_{t_1} \, \int_{0}^{t_1}ds \hat{A}_{s} \nonumber \\
&&+\int_{t_1}^{t}ds^{\prime}\int_{t_1}^{s^{\prime}}ds \hat{A}_{s^{\prime}} \hat{A}_{s}\,  \hat{E}_{t_1}\Big],
\end{eqnarray}
\begin{eqnarray}
\hat{R}_{t\vert t_{1}}(3)&=&\sqrt{dt_1}\, \hat{T}_t \, \Big[ \xi_{t_1}\int_{0}^{t_1}ds^{\prime}\int_{0}^{s^{\prime}}ds\hat{A}_{s^{\prime}} \hat{A}_{s}\nonumber\\ 
&&+\xi_{t_1}\int_{t_1}^{t}ds^{\prime}\int_{0}^{t_1}ds\hat{A}_{s^{\prime}} \hat{A}_{s}+\xi_{t_1}\int_{t_1}^{t}ds^{\prime}\int_{t_1}^{s^{\prime}}ds\hat{A}_{s^{\prime}} \hat{A}_{s} \nonumber\\
&&+\hat{E}_{t_1}\int_{0}^{t_1}ds^{\prime\prime}\int_{0}^{s^{\prime\prime}}ds^{\prime}\int_{0}^{s^{\prime}}ds \hat{A}_{s^{\prime\prime}}\hat{A}_{s^{\prime}}\hat{A}_{s}\nonumber\\
&&+\int_{t_1}^{t}ds\hat{A}_{s}\hat{E}_{t_1}\int_{0}^{t_1}ds^{\prime\prime}\int_{0}^{s^{\prime\prime}}ds^{\prime} \hat{A}_{s^{\prime\prime}}\hat{A}_{s^{\prime}}\nonumber\\
&&+\int_{t_1}^{t}ds^{\prime\prime}\int_{t_1}^{s^{\prime\prime}}ds^{\prime} \hat{A}_{s^{\prime\prime}}\hat{A}_{s^{\prime}}
\hat{E}_{t_1}\int_{0}^{t_1}ds\hat{A}_{s}\nonumber\\
&&+\int_{t_1}^{t}ds^{\prime\prime}\int_{t_1}^{s^{\prime\prime}}ds^{\prime}\int_{t_1}^{s^{\prime}}ds \hat{A}_{s^{\prime\prime}}\hat{A}_{s^{\prime}}\hat{A}_{s}\hat{E}_{t_1} \Big],
\end{eqnarray}
and so on.
\end{enumerate}

The {\it a priori} state of $\mathcal{S}$ at time $t$ can be express by the conditional operators as 
\begin{equation}\label{sol7}
\sigma_{t}=\rho_{t\vert 0}+\sum_{s=1}^{+\infty}\int_{0}^{t}dt_{s}\int_{0}^{t_{s}}dt_{s-1}\ldots
\int_{0}^{t_{2}}dt_{1}\rho_{t\vert t_{s},t_{s-1},\ldots,t_{2},t_{1}}
\end{equation}
where
\begin{eqnarray}
\rho_{t\vert 0}=\sum_{m=0}^{+\infty}u_{t}^{m}\ket*{\psi^{n}_{t\vert 0}(m)}\bra*{\psi^{n}_{t\vert 0}(m)}
\end{eqnarray}
with
\begin{equation}
u_{t}=\int_{t}^{+\infty}dt^{\prime}\lvert \xi_{t^{\prime}}\rvert^2
\end{equation}
and
\begin{eqnarray}
\lefteqn{dt_{s}dt_{s-1}\ldots dt_{1}\rho_{t\vert t_{s},t_{s-1},\ldots,t_{2},t_{1}}=}
\\&&\sum_{m=0}^{+\infty}u_{t}^{m}\ket*{\psi^{n}_{t\vert t_{s},t_{s-1},\ldots,t_{2},t_{1}}(m)}\bra*{\psi^{n}_{t\vert t_{s},t_{s-1}\ldots,t_{2},t_{1}}(m)}\nonumber.
\end{eqnarray}
We have here the sum over all photons detection pathways that might take place from $0$ to time $t$. They could involve $s$ detections where $s$ could change from $0$ to $\infty$. The operators under integrals are interpreted as the unnormalized conditioned density operator of $\mathcal{S}$ associated with different scenarios of photon detections. For instance, $\rho_{t\vert 0}$ refers to the situation when we do not observe any photons up to $t$ while $\rho_{t\vert t_{s},t_{s-1},\ldots,t_{2},t_{1}}$ to the case when $s$ photons were registered in the intervals $[t_{1},t_{1}+dt_{1})$, $[t_{2},t_{2}+dt_{2})$, $\ldots$, $[t_{s},t_{s}+dt_{s})$, where $t_1 < t_{2}<\ldots < t_s$. The formula (\ref{sol7}) is a decomposition of the reduced state of $\mathcal{S}$ by means of quantum trajectories associated with the counting process $N_t$. We would like to emphasis that this solution is not unique i.e. considering homodyne or heterodyne measurement schemes we would get different stochastic representation of $\sigma_t$.   

One can use the conditional vectors to find the statistics of the counting of the output photons. The probability of not observing any photons up to time $t$ is given as
\begin{equation}
P_{0}^{t}(0)=\sum_{m=0}^{+\infty}\bra*{\psi^{n}_{t\vert 0}(m)}
\ket*{\psi^{n}_{t\vert 0}(m)}\left(\int_{t}^{+\infty}dt^{\prime}\lvert \xi_{t^{\prime}}\rvert ^2\right)^{m}.
\end{equation}
The exclusive probability density $ p_{0}^{t}(t_{s}, t_{s-1}, \ldots, t_{2}, t_{1})$ for a trajectory corresponding to $s$ detections in the interval from $0$ to $t$ in the intervals $[t_{1},t_{1}+dt_{1})$, $[t_{2},t_{2}+dt_{2})$, $\ldots$, $[t_{s},t_{s}+dt_{s})$, where $0<t_1 < t_{2}<\ldots < t_s$ is given as
\begin{eqnarray}
\lefteqn{ p_{0}^{t}(t_{s}, t_{s-1}, \ldots, t_{2}, t_{1}){dt_{s}dt_{s-1}\ldots dt_{1}} =}
\\&& \sum_{m=0}^{+\infty}\bra*{\psi_{t\vert t_{s}, t_{s-1}, \ldots, t_{2}, t_{1}}^{n}(m)}\ket*{\psi_{t\vert t_{s}, t_{s-1}, \ldots, t_{2}, t_{1}}^{n}(m)} \left(\int_{t}^{+\infty}dt^{\prime}\rvert \xi_{t^{\prime}}\vert^2\right)^{m}\nonumber
\end{eqnarray}
Hence the probability of registering exactly $s$ photons up to time $t$ is
\begin{equation}
P_{0}^{t}(s)\!=\!\int_{0}^{t}\!dt_{s}\!\int_{0}^{t_{s}}\!dt_{s-1}\!\ldots\!\int_{0}^{t_{2}}\!dt_{1}
p_{0}^{t}(t_{s}, \!t_{s-1}, \!\ldots \!, \!t_{2}, t_{1}) .
\end{equation}

\section{Example: the photon profile for the most efficient transfer of the field photons into the cavity}

Let us assume that a quantum system interacting with a wave packet is a cavity mode---a harmonic oscillator.  
In this Section, we study the problem of choosing an optimal photon profile that gives the most efficient transfer of photons from the wave packet into the cavity. First, we assume that the traveling field is prepared in the number state, and then we consider the case of the input field in the squeezed number state. We write down the Hamiltonian of the system in the rotating frame, that is 
\begin{equation}
\hat{H}_{\mathcal{S}}=\Delta \hat{a}^{\dagger}\hat{a},
\end{equation}
where $\Delta=\omega_{0}-\omega_{c}$, where $\omega_{0}$ is the frequency of the cavity mode and $\omega_{c}$ stands for the central frequency of the input wave packet.
We take the coupling operator of the form
\begin{equation}
\hat{L}=\sqrt{\Gamma}\hat{a},\;\;
\end{equation} 
where $\Gamma>0$. Let us emphasis that $\xi_{t}$ is a slowly-varying envelop of the pulse \cite{Gross18,Garrison2008}. 
Let us assume that the harmonic oscillator is initially in the vacuum state $\ket{0}$.  
The probability that there are $n$ photons inside cavity at time $t$ is given by $P_{n_{\xi}}(t)=\|\ket*{\psi^{n}_{t\vert 0}(0)}\|^2$. The conditional vector $\ket*{\psi^{n}_{t\vert 0}(0)}$ refers to the case when we do not detect any photons up to $t$ and the input field after $t$ is in the vacuum state. Clearly, it means that all photons from the wave packet were absorbed by the cavity and stayed there up to $t$. One can check that then
	\begin{eqnarray}
	P_{n_{\xi}}(t) &=&\Gamma^{n}n!e^{-n\Gamma t} \left\vert \int_{t_{0}}^{t}dt_{n}\int_{t_{0}}^{t_n}dt_{n-1} \right.\nonumber\\&&
	\left.\ldots\int_{t_{0}}^{t_{3}}dt_{2}\int_{t_0}^{t_{2}}
	dt_{1} \prod_{i=1}^{n}\xi_{t_i}e^{\left(i\Delta+\frac{\Gamma}{2}\right)t_{i}}\right\vert^2,
	\end{eqnarray}
	which can be expressed as
	\begin{equation}\label{pn}
	P_{n_{\xi}}(t) =\Gamma^{n}e^{-n\Gamma t}\left\vert\int_{t_{0}}^{t}ds\xi_{s}e^{\left(i\Delta+\frac{\Gamma}{2}\right)s}\right\vert^{2n}.
	\end{equation}
	We have assumed here that the interaction between the systems starts in an arbitrary moment $t_0$.
\begin{Proposition}\label{pro1}	
	The maximum value of the probability of $n$ excitations at time $t>t_{0}$ for the cavity mode prepared in the vacuum state and the input field in $\ket*{n_{\xi}}$ reads as
	\begin{equation}\label{A-max}
	P^{\rm max}_{n_{\xi}}(t) :=   \max_{\xi} P_{n_{\xi}}(t) =e^{-n\Gamma t}\left(e^{\Gamma t}-e^{\Gamma t_0}\right)^{n},
	\end{equation}
	and is realized only at the resonance (i.e. $\Delta =0$)  by the pulse of the profile
	\begin{equation}\label{xi-max}
	\xi_{s} = \sqrt{\frac{\Gamma}{e^{\Gamma t}-e^{\Gamma t_0}}} e^{\frac{\Gamma}{2}s} 
	\end{equation}
	for $s \in[t_0,t]$, and $\xi_{s}=0$ elsewhere. 
\end{Proposition}
Thus we obtain a perfect transfer, i.e. $P_{n}(t)=1$, for an exponential pulse rising in the interval $(-\infty,t]$. If the input field is prepared in the squeezed number state $\ket*{n_{\gamma,\xi}}$, then the probability that the mean number of photons inside the cavity at time $t$ is equal to $c^2n+s^2(n+1)$ is given by 
\begin{equation}\label{pngamma}
P_{n_{\gamma,\xi}}(t) =\sum_{k=0}^{+\infty}\Gamma^{k}k!e^{-k\Gamma t}\left\vert\int_{t_{0}}^{t}ds\xi_{s}e^{\left(i\Delta+\frac{\Gamma}{2}\right)s}\right\vert^{2k}\lvert a_{k}(n_{\gamma})\rvert^2.
\end{equation}

\begin{Proposition}\label{pro2}
	The maximum value of the probability of the transfer of the wave packet photons into the cavity at time $t>t_{0}$ for the cavity mode prepared in the vacuum state and the input field in $\ket*{n_{\gamma,\xi}}$ is
	\begin{equation}
	P^{\rm max}_{n_{\gamma,\xi}}(t) :=   \max_{\xi} P_{n_{\gamma,\xi}}(t) =\sum_{k=0}^{+\infty}e^{-k\Gamma t}\left(e^{\Gamma t}-e^{\Gamma t_0}\right)^{k}\lvert a_{k}(n_{\gamma})\rvert^2,
	\end{equation}
	and is realized only at the resonance (i.e. $\Delta =0$)  by the pulse of the profile
	\begin{equation}
	\xi_{s} = \sqrt{\frac{\Gamma}{e^{\Gamma t}-e^{\Gamma t_0}}} e^{\frac{\Gamma}{2}s} 
	\end{equation}
	for $s \in[t_0,t]$, and $\xi_{s}=0$ elsewhere. 
\end{Proposition}
The proofs for (\ref{pro1}) and (\ref{pro2}) are given in \ref{D}.

\section{Summary}

We have derived the sets of filtering and master equations for an open 
quantum system coupled to the continuous-mode field prepared in the squeezed number state. 
We have shown that the quantum system becomes then entangled with the input and output parts of the field and the conditional and unconditional evolutions of the system are given by infinite sets of coupled equations. We have determined the evolution of the quantum system starting from the discrete in-time model where the traveling field is defined by a sequence of harmonic oscillators that interact one by one with the quantum system. The harmonic oscillators are subsequently monitored. The procedure of determining the conditional state of the system for discrete measurements is based on von Neumann's projection postulate. Clearly, random results of the measurements lead to random sequence of the system states. The discrete model gives an intuitive picture to interaction of the wave packet in the squuezed state with the quantum system and to the stochastic evolution. 
 We have finally obtained the results for the continuous in-time measurement and evolution of the system. Our sets of differential stochastic equations and master equations agree with the results determined by means of QSC in \cite{Gross2022}. We have not only determined the filtering equations but we have also found the analytical formulae for quantum trajectories associated with the counting measurement performed on the output field and we have given the physical interpretation to the quantum trajectories.  
 In this paper one can find, moreover, a decomposition of the {\it a priori} state by means of conditional operators associated with the stochastic counting process. It is a generalization of the famous formula for unraveling of the reduced evolution of an open quantum system coupled to the vacuum field \cite{Car93}.  The conditional operators were also applied to obtain the formula for the probability of zero counts up to $t$ and the expressions for exclusive probability densities which allow to fully characterize the photon statistics in the output field. In simple illustrative example we have shown how to use the conditional vectors to solve the problem of the transfer of the photons from the wave packet to the cavity. By choosing the rising profile for the photons the probability of a perfect transfer could be arbitrary close to unity.    


\appendix

\section{Proof of Theorem (\ref{TH-1})} \label{B}

We proof Theorem (\ref{TH-1}) using an induction. Thus we will show that if (\ref{condst2}) holds for $j$, then it also holds for $j+1$. We start from an observation that the conditional vector $\ket*{\Psi_{j\vert  \pmb{\eta}_j}}$ from $\mathcal{H}^{[j}_{\mathcal{E}}\otimes \mathcal{H_{S}}$ can be rewritten in the form
\begin{equation}
\ket*{\Psi_{j\vert \pmb{\eta}_j}^{n}}=
\sum_{m=0}^{+\infty}\sum_{m^{\prime}=0}^{m}\sqrt{\binom{m}{m^{\prime}}} (\sqrt{\tau} \xi_{j})^{m^{\prime}} \ket*{m^{\prime}}_{j}\otimes\ket*{(m-m^{\prime})_{\xi}}_{[j+1}\otimes\ket*{\psi_{j\vert\pmb{\eta}_j}^{n}(m)}
\end{equation}
which follows from the fact that
\begin{equation}\label{}
\ket*{m_{\xi}}_{[j+1} = \sum_{m^{\prime}=0}^{m}\sqrt{\binom{m}{m^{\prime}}} (\sqrt{\tau} \xi_{j})^{m^{\prime}} \ket*{m^{\prime}}_{j}\otimes \ket*{(m-m^{\prime})_{\xi}}_{[j+1}.
\end{equation}
From an action of the unitary operator $\hat{\mathbb{V}}_{[k}$ on  $\ket*{\Psi_{j\vert \pmb{\eta}_j}}$ we have
\begin{eqnarray}
\lefteqn{\hat{\mathbb{V}}_{[k} \,\ket*{\Psi_{j\vert \pmb{\eta}_j}^{n}} =} \nonumber\\ &=& \sum_{m=0}^{+\infty} \sum_{m^{\prime}=0}^{m} \sqrt{\binom{m}{m^{\prime}}} (\sqrt{\tau} \xi_j)^{m^{\prime}} \sum_{i=0}^{+\infty} \ket{i}_{j} \otimes \ket*{(m-m^{\prime})_\xi}_{[j+1} \otimes V_{im^{\prime}} \ket*{\psi_{j\vert \pmb{\eta}_j}^{n}(m)}.
\end{eqnarray}

The conditional vector $\ket*{\Psi_{j+1\vert  \pmb{\eta}_{j+1}}^{n}}$ from $\mathcal{H}_{\mathcal{E}}^{[j+1}\otimes \mathcal{H}_{S}$ is defined by
\begin{equation}\label{measurement}
\left(\ket*{\eta_{j+1}}_{j}\bra*{\eta_{j+1}} \otimes\mathbbm{1}_{\mathcal{E}}^{[j+1}\otimes\mathbbm{1}_{S}\right) \,\hat{\mathbb{V}}_{[k}\,\ket*{\Psi_{j\vert  \pmb{\eta}_j}^{n}}=\ket*{\eta_{j+1}}_{j}\otimes\ket*{\Psi_{j+1\vert  \pmb{\eta}_{j+1}}^{n}},
\end{equation}
where $\eta_{j+1}\in \mathbb{N}$ is the random result of the measurement of (\ref{obs}) at the time $\tau(j+1)$. Hence we obtain the formula 
\begin{eqnarray}
\lefteqn{\ket*{\Psi_{j+1\vert \pmb{\eta}_{j+1}}^{n}}=} \nonumber \\ &=& \sum_{m=0}^{+\infty} \sum_{m^{\prime}=0}^{m} \sqrt{\binom{m}{m^{\prime}}} (\sqrt{\tau} \xi_j)^{m^{\prime}} \ket*{(m-m^{\prime})_\xi}_{[j+1} \otimes V_{\eta_{j+1}m^{\prime}} \ket*{\psi_{j\vert \pmb{\eta}_j}^{n}(m)}.
\end{eqnarray}
To get the recurrence formulae (\ref{rec}) we change the index of summation 
by inserting $s=m-m^{\prime}$ such that
\begin{eqnarray}\label{cond2}
\ket*{\Psi_{j+1\vert \pmb{\eta}_{j+1}}} = \sum_{s=0}^{+\infty}  \ket*{s_\xi}_{[j+1} \otimes \sum_{m^{\prime}=0}^{+\infty} \sqrt{\binom{s+m^{\prime}}{m^{\prime}}} (\sqrt{\tau} \xi_j)^{m^{\prime}} V_{\eta_{j+1}m^{\prime}} \ket*{\psi_{j\vert \pmb{\eta}_j}^{s+m^{\prime}}}.
\end{eqnarray}
And this ends the proof. \hfill $\Box$

\section{Master equations}\label{C} 

In order to derive the set of master equations describing the evolution of the system $\mathcal{S}$ we can use the representation of the squeezed number states $\{\ket*{n_{\gamma,\xi}}\}$ in the basis of the photon number states $\{\ket*{m_{\xi}}\}$ and the fact that evolution operator is linear. Thus, for the system operator  
\begin{equation}
\sigma_{j}^{n^{\prime},n^{\prime\prime}}= \mathrm{Tr}_{\mathcal{E}}(\hat{U}_{j}
\ket*{n^{\prime}_{\gamma,\xi}}\bra*{ n^{\prime\prime}_{\gamma,\xi}}
\otimes \ket*{\psi_{0}}\bra*{\psi_{0}}\hat{U}_{j}^{\dagger}). 
\end{equation}
by the expansion 
\begin{equation}
\ket*{n^{\prime}_{\gamma,\xi}}\bra*{ n^{\prime\prime}_{\gamma,\xi}}=\sum_{m^{\prime}, m^{\prime\prime}=0}^{+\infty}a_{m^{\prime}}(n^{\prime}_{\gamma})a_{m^{\prime\prime}}^{\ast}(n^{\prime\prime}_{\gamma})\ket*{m^{\prime}_{\xi}}\bra*{ m^{\prime\prime}_{\xi}}
\end{equation}
we obtain
\begin{equation}\label{representation}
\sigma_{j}^{n^{\prime},n^{\prime\prime}}= \sum_{m^{\prime},m^{\prime\prime}=0}^{+\infty}a_{m^{\prime}}(n^{\prime}_{\gamma})a_{m^{\prime\prime}}(n^{\prime\prime}_{\gamma})\varrho_{j}^{m^{\prime}m^{\prime\prime}},
\end{equation}
where 
\begin{equation}
\varrho_{j}^{m^{\prime},m^{\prime\prime}}= \mathrm{Tr}_{\mathcal{E}}(\hat{U}_{j}\ket*{m_{\xi}^{\prime}}\bra*{ m_{\xi}^{\prime\prime}}\otimes \ket*{\psi_{0}}\bra*{\psi_{0}}\hat{U}_{j}^{\dagger}). 
\end{equation}
The {\it a priori} state of $\mathcal{S}$ for the initial state of the composed system defined by (\ref{ini0}) is given by 
\begin{equation}\label{sigma}
\sigma_{j}^{n,n}=\sum_{m^{\prime},m^{\prime\prime}=0}^{+\infty}a_{m^{\prime}}(n_{\gamma})a_{m^{\prime\prime}}^{\ast}(n_{\gamma})\varrho_{j}^{m^{\prime},m^{\prime\prime}}. 
\end{equation}
It was shown in \cite{Baragiola12,Baragiola17,Dabrowska19} that the system operator $\varrho_{j}^{m^{\prime},m^{\prime\prime}}$ satisfy the equation 
\begin{eqnarray}\label{master3}
{\varrho}_{j+1}^{m^{\prime},m^{\prime\prime}}&=&{\varrho}_{j}^{m^{\prime},m^{\prime\prime}}+\mathcal{L}\varrho_{j}^{m^{\prime},m^{\prime\prime}}+ \sqrt{m^{\prime}}[{\varrho}_{j}^{m^{\prime}-1,m^{\prime\prime}},\hat{L}^{\dagger}]\xi_{j}
\nonumber\\&&+[\hat{L}, {\varrho}_{j}^{m^{\prime},m^{\prime\prime}-1}]\sqrt{m^{\prime\prime}}\xi^{\ast}_{j},  
\end{eqnarray}
where $\mathcal{L}$ is the superoperator in the form of (\ref{supero}). We have here an infinite number of coupled equations. 
Making use of (\ref{rel}) we obtain
\begin{eqnarray}
\sum_{m^{\prime}=1}^{+\infty}\sum_{m^{\prime\prime}=0}^{+\infty}\sqrt{m^{\prime}}a_{m^{\prime}}(n^{\prime}_{\gamma})a_{m^{\prime\prime}}^{\ast}(n^{\prime\prime}_{\gamma})\varrho_{j}^{m^{\prime-1},m^{\prime\prime}}\nonumber\\=\sum_{m^{\prime},m^{\prime\prime}=0}^{+\infty}\sqrt{m^{\prime}+1}a_{m^{\prime}+1}(n^{\prime}_{\gamma})a_{m^{\prime\prime}}^{\ast}(n^{\prime\prime}_{\gamma})\varrho_{j}^{m^{\prime},m^{\prime\prime}}\nonumber\\=
\sqrt{n^{\prime}}c\sigma_{j}^{n^{\prime}-1,n^{\prime\prime}}-\sqrt{n^{\prime}+1}se^{2i\phi}\sigma_{j}^{n^{\prime}+1,n^{\prime\prime}}. 
\end{eqnarray}
Using this result one can easily check that from (\ref{master3}) we obtain for the operator $\sigma_{t}^{n^{\prime},n^{\prime\prime}}$ the master equation (\ref{master1}). 

It is worth to notice that one can use the photon number representation and the set of equations (\ref{master3}) to find approximate solution defining the reduced state  of $\mathcal{S}$. It is clear that according to the convergence of (\ref{sigma}) the contribution to (\ref{representation}) of expressions for $m^{\prime},m^{\prime}\to \infty$ goes to zero.

\section{Proof to \ref{pro1} and \ref{pro2}}\label{D} 
 
 Let us introduce $\tilde{\xi}_{s}=e^{-i\omega_{c}s}\xi_{s}\in\mathbb{C}$. To maximize the probabilities (\ref{pn}) and (\ref{pngamma}) one has to maximize the expression
 \begin{equation}\label{}
 \lvert\int_{t_0}^{t}ds \tilde{\xi}_{s}e^{\left(i\omega_{0}+\frac{\Gamma}{2}\right) s}\rvert^2 = \lvert\bra{\tilde{\xi}}\ket{f}_{t_0,t} \rvert^2
 \end{equation}
 where $f_{s} =  e^{\left(-i\omega_{0}+\frac{\Gamma}{2}\right) s}$, and we have introduced an inner product
 \begin{equation}\label{ip}
 \bra{\tilde{\xi}}\ket{f}_{t_0,t} := \int_{t_0}^t\tilde{\xi}_{s} f^{\ast}_{s} ds .
 \end{equation}
The maximum value of (\ref{ip}) we obtain for the profile $\tilde{\xi}_{s}$ parallel to $f_{s}$, that is, $\tilde{\xi}_{s} = c(t_0,t) f_{s}$, where $c(t_{0},t)$ is a constant (depending on fixed times $t_{0}$ and $t$), such that
 \begin{equation}\label{}
 \bra{ \tilde{\xi}}\ket{\tilde{\xi}} = 1 .
 \end{equation}
 These conditions allows to calculate
 \begin{equation}
 \tilde{\xi}_{s} = \sqrt{ \frac{\Gamma}{e^{\Gamma t}-e^{\Gamma t_0}} } \, e^{\left(-i\omega_{0}+\frac{\Gamma}{2}\right)s}
 \end{equation}
 for $s \in[t_0,t]$, and ${\xi}_{s}=0$ for $s > t$ and $s<t_{0}$. Clearly, the maximum excitation is realized at resonance  $\omega_{c}=\omega_{0}$, i.e. $\Delta_{0}=0$. \hfill $\Box$

\section*{Acknowledgments}

This research was supported by the National Science Centre project 2018/30/A/ST2/00837.





\end{document}